%% file: main.tex
\pdfoutput=1
\documentclass{article}

\PassOptionsToPackage{table}{xcolor}

\usepackage[preprint]{conference}

\usepackage{hyperref}

\usepackage{amsmath}
\usepackage{amssymb}
\usepackage{mathtools}
\usepackage{amsthm}
\usepackage[capitalize,noabbrev]{cleveref}

\usepackage{amsfonts}
\usepackage{nicefrac}
\usepackage{microtype}
\usepackage{graphicx}
\usepackage{subcaption}
\usepackage{booktabs}
\usepackage{changepage}
\usepackage{outlines}
\usepackage{multirow}
\usepackage{tabularx}
\usepackage{enumitem}
\usepackage{placeins}
\usepackage{listings}
\usepackage{tikz}
\usetikzlibrary{shapes, arrows.meta, positioning, fit, backgrounds, calc, decorations.pathreplacing}
\definecolor{codegreen}{rgb}{0,0.6,0}
\definecolor{codegray}{rgb}{0.5,0.5,0.5}
\definecolor{codepurple}{rgb}{0.58,0,0.82}
\definecolor{backcolour}{rgb}{0.95,0.95,0.92}
\definecolor{myblue}{RGB}{70, 130, 180}
\definecolor{mygreen}{RGB}{60, 179, 113}
\definecolor{myred}{RGB}{205, 92, 92}
\definecolor{mygray}{RGB}{100, 100, 100}
\definecolor{lightgray}{RGB}{245, 245, 245}
\definecolor{mylinkblue}{rgb}{0,0.08,0.45}
\hypersetup{
  colorlinks=true,
  linkcolor=mylinkblue,
  citecolor=mylinkblue,
  urlcolor=mylinkblue,
  filecolor=mylinkblue,
  pdfborder=0 0 0,
}
\lstdefinestyle{mystyle}{
    backgroundcolor=\color{backcolour},   
    commentstyle=\color{codegreen},
    keywordstyle=\color{magenta},
    numberstyle=\tiny\color{codegray},
    stringstyle=\color{codepurple},
    breakatwhitespace=false,         
    breaklines=true,                 
    keepspaces=true,                 
    numbers=left,                    
    numbersep=5pt,                  
    showspaces=false,                
    showstringspaces=false,
    showtabs=false,
    tabsize=2,
    xleftmargin=.05\textwidth,
    xrightmargin=.05\textwidth
}

\usepackage[disable,colorinlistoftodos,textsize=tiny]{todonotes}

\newcommand{\candidate}[1]{c_{#1}}

\newcommand{\generator}{\pi_{\textrm{G}}}

\newcommand{\orm}{\ormVerifier{}}

\newcommand{\verifier}[1]{V_{#1}}
\newcommand{\strongestVerifier}{\verifier{\text{all}}}
\newcommand{\weakestVerifier}[1]{\verifier{c_{#1}}}
\newcommand{\weakVerifier}[1]{\verifier{S_{#1}}}
\newcommand{\ormVerifier}{\verifier{\theta}}
\newcommand{\pruneOrmVerifier}[1]{\verifier{\text{#1}+\theta}}

\newcommand{\tenTestVerifier}{\verifier{S_{10}+\theta}}
\newcommand{\nTestPruneVerifier}{\verifier{S_{N}+\theta}}

\newcommand{\bestofk}[1]{\textrm{Best-of-}\ifx\relax#1\relax k\else#1\fi}

\newcommand{\numCoresUsed}{32}

\newcommand{\trainNumSampled}{128}

\newcommand{\lintVerifier}{\verifier{\text{Lint}}}

\definecolor{textred}{HTML}{b88282}
\definecolor{textgreen}{HTML}{82b891}

\newcommand{\ormFasterPct}{8.60$\times$}

\newcommand{\ormNaiveAccPct}{32.86\%}

\newcommand{\tenTestPPSStrongImprove}{11.64$\times$}
\newcommand{\tenTestAccStrongWorse}{8.26\%}

\NewExpandableDocumentCommand\mcc{O{1}m}{\multicolumn{#1}{c}{#2}}

\conftitlerunning{Pareto Optimal Code Generation}

\begin{document}

\twocolumn[
\conftitle{Pareto Optimal Code Generation}

\begin{confauthorlist}
\confauthor{Gabriel Orlanski}{wisc}
\confauthor{Nicholas Roberts}{wisc}
\confauthor{Aws Albarghouthi}{wisc}
\confauthor{Frederic Sala}{wisc}
\end{confauthorlist}

\confaffiliation{wisc}{University of Wisconsin-Madison}

\confcorrespondingauthor{Gabriel Orlanski}{gorlanski@cs.wisc.edu}

\vskip 0.3in
]

\printAffiliationsAndNotice{}

\begin{abstract}
Generate-then-rank is the dominant test-time scaling (TTS) paradigm for code generation, but scaling accuracy by sampling and executing more candidates makes comprehensive verification a major computational bottleneck. 
This creates an inherent trade-off between accuracy and compute that, despite its importance to TTS, is often ignored.
Specifically, faster but noisier signals, such as outcome reward models (ORMs), are dismissed as suboptimal.
We frame verifier selection as a Pareto optimization problem and empirically map the accuracy-throughput frontier across signals, including the full test suite, heuristics for selective execution, and ORMs, across four Python benchmarks. 
We show that ORMs are most effective at optimizing the Pareto curve when pruning is used in the generate-then-rank pipeline--known as staged verification--where lightweight filters remove obviously incorrect solutions, including candidates with small syntactic or character-level bugs, before expensive verification. 
Our pruning analysis shows that eliminating incorrect yet highly ranked candidates (often character-level bugs) prevents wasted compute on incorrect tokens.
We find that ORMs with staged verification shift the Pareto frontier outward, achieving 11.64x higher throughput at a cost of 8.26\% accuracy relative to full test-suite verification.
\end{abstract}

\input{sections/introduction}

\input{sections/preliminaries}

\input{sections/research_questions}

\input{sections/results}

\input{sections/related_work}
\input{sections/conclusion}


\section*{Impact Statement}
Staged verification reduces compute costs for large-scale code generation by pruning incorrect candidates before expensive validation. However, treating lightweight filters as a substitute for thorough testing could increase the risk of deploying incorrect or insecure code, and lowering verification costs may amplify harmful uses of high-volume generation. We emphasize that staged verification is a selection tool, not a replacement for full test suites. Sandboxed execution and human review remain essential before deployment.

\bibliography{references}
\bibliographystyle{conference}
\newpage
\FloatBarrier
\appendix
\onecolumn
\input{sections/appendix/train_data.tex}

\input{sections/appendix/example_program.tex}
\input{sections/appendix/bug_examples.tex}
\input{sections/appendix/prompts.tex}
\input{sections/appendix/execution_details}
\input{sections/appendix/data.tex}

\input{sections/appendix/eval_data.tex}
\input{sections/appendix/scale_graphs.tex}
\input{tables/results_std_table.tex}
\input{sections/appendix/tables.tex}
\end{document}

%% file: sections/introduction.tex
\section{Introduction}
\begin{figure*}
  \centering
  \includegraphics[width=\textwidth]{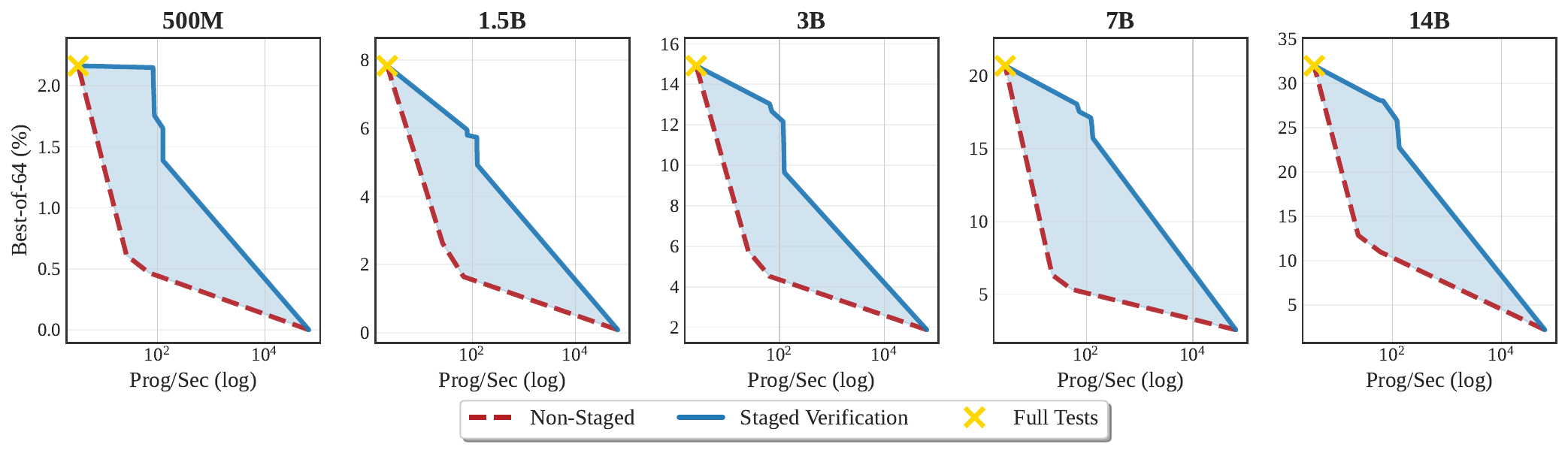}
  \caption{\textbf{Staged Verification Shifts the Pareto Frontier.} Each panel shows one generator size (500M to 14B) on CodeContests. \textcolor[HTML]{b41f24}{Red dashed}: non-staged verification. \textcolor[HTML]{1f77b4}{Blue solid}: staged verification, which filters before ORM ranking. \textcolor[HTML]{FFD700}{Gold $\times$}: full test suite. The shaded region shows gains from staging; the frontier shifts outward across all generator sizes.}
  \label{fig:overview}
\end{figure*}

Generate-then-rank has become the dominant paradigm for test-time scaling in code generation. The pipeline \emph{generates} many candidates from a model, \emph{verifies} each one against a test suite, and \emph{returns} the highest-ranked correct solution. With hundreds or thousands of candidates, verification becomes the bottleneck. This is especially true for more comprehensive SWE-Bench~\citep{jimenez_swe-bench_2024} test suites, which cost orders of magnitude more than the much lighter-weight HumanEval~\citep{chen_evaluating_2021} unit tests.  

Recent work discards neural reward models when reliable test suites exist~\citep{deepseek-ai_deepseek-r1_2025,muennighoff_s1_2025,lambert_tulu_2025}, but this assumes verification cost is negligible relative to generation cost. As test suites grow more complex, this assumption breaks down.

This cost creates a fundamental tradeoff. Full test suites maximize correctness but bottleneck throughput; lightweight checks like compilation and partial execution maximize throughput but miss bugs. Outcome Reward Models (ORMs) offer a third option: neural rankers that score candidates in a single pass, with inference cost depending on candidate length rather than task complexity. These three approaches define a Pareto frontier of accuracy vs. throughput. However, where ORMs sit on this frontier relative to execution-based methods---and whether combining the two shifts the frontier outward---has not been studied. 

The question is complicated by an asymmetry. Execution-based tradeoffs are \emph{transparent}: using fewer tests predictably misses more bugs. ORM tradeoffs are \emph{opaque}: these models can assign high scores to candidates with obvious defects, including arithmetic mistakes and logical errors in intermediate reasoning steps~\citep{lightman_lets_2023,wang_math-shepherd_2024}. Cheap execution filters catch precisely these failure modes. Prior work evaluates verifiers in isolation, reporting accuracy or throughput but not both, leaving the frontier unmapped~\citep{inala_fault-aware_2022,zhang_coder_2022,ni_lever_2023,cobbe_training_2021,snell_scaling_2024}.

\textbf{We empirically map the accuracy-throughput frontier for code verification.} We focus on single-file Python benchmarks where verification cost already varies substantially; repository-scale tasks like SWE-Bench are future work. Our investigation spans three verification signals across four benchmarks: full test suites, partial test execution, and ORMs. We evaluate both standard generate-then-rank and \textit{staged verification}, where lightweight filters remove obvious failures before ORM ranking (\autoref{fig:pipeline}). \textbf{Staged verification dominates naive ORM ranking across much of the frontier: filtering eliminates candidates with syntactic bugs that ORMs systematically over-rank, preventing wasted compute on incorrect solutions} (\autoref{fig:overview}).

Our contributions are as follows:
\begin{itemize}[noitemsep]
  \item \textbf{We map the accuracy-throughput frontier:} ORMs are \ormFasterPct{} faster than full test suites and \ormNaiveAccPct{} more accurate than majority voting.
  \item \textbf{We show staged verification shifts the frontier outward:} Filtering before ranking is \tenTestPPSStrongImprove{} faster and \tenTestAccStrongWorse{} less accurate than full test suites.
  \item \textbf{We characterize ORM failure modes and demonstrate that staged verification addresses them:} Gains stem from removing incorrect candidates that ORMs over-rank, particularly code with small syntactic bugs that preserve overall structure.
  \item \textbf{We show staged verification does not require careful test curation:} Random subsets of tests achieve false positive rates comparable to those of curated subsets, with low variance across selections.
\end{itemize}

We have also released code reproducing our experiments.\footnote{\url{https://github.com/SprocketLab/orm-code-verifier}}

%% file: sections/preliminaries.tex
\section{The Verification Pareto Frontier}

This section formalizes the verification problem. We first define the metrics that characterize the Pareto frontier (\autoref{sec:problem-setup}), then introduce staged verification as a strategy for combining cheap filters with neural rankers (\autoref{sec:staged-verification}), and finally describe ORMs trained with the Bradley-Terry objective as fixed-cost verifiers (\autoref{sec:training-orm}).

Verification strategies occupy different points on an accuracy-throughput Pareto frontier. Strong verifiers (full test suites) maximize accuracy but bottleneck throughput; weak verifiers (syntax checks, partial execution) maximize throughput but sacrifice accuracy. ORMs offer fixed inference cost regardless of test suite complexity, but where they sit on this frontier—and whether combining them with lightweight filters can shift the frontier outward—remains unexplored. We focus on Python benchmarks with test-based verification; repository-scale tasks are future work.

\begin{figure*}[tb]
    \centering
    \includegraphics[width=1.0\textwidth]{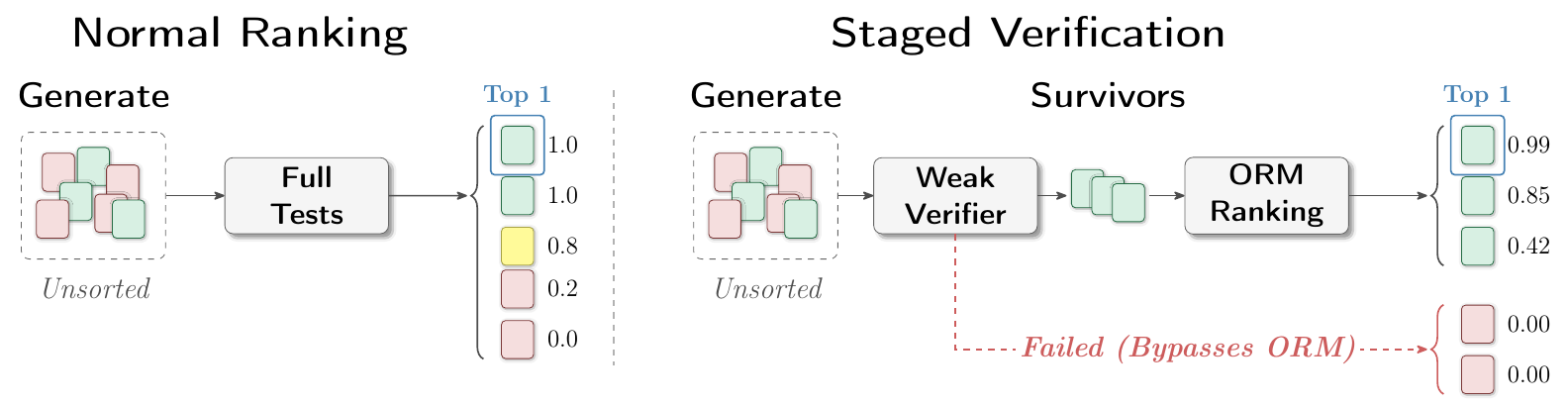}
    \caption{Comparison of verification strategies. \textbf{Normal ranking} runs all tests to produce a high-quality ordering but is slow, especially with a large test suite. \textbf{Staged verification} uses a weak verifier to filter obvious failures, then ranks survivors using an ORM.}
    \label{fig:pipeline}
\end{figure*}

\subsection{Problem Setup}\label{sec:problem-setup}

Given a programming problem $x$, a generator produces candidates $C = \{c_1, \ldots, c_n\}$. Each candidate is scored by a verifier $\verifier{}$, and the top-ranked candidate is returned. We assume each problem is associated with test cases that determine the correctness of a candidate, as is standard in competitive programming~\citep{li_competition-level_2022,jain_livecodebench_2024} and software engineering benchmarks~\citep{jimenez_swe-bench_2024}.

\paragraph{Accuracy.} We measure ranking quality using Best-of-$k$~\citep[Bof$k$;][]{hosseini_v-star_2024}: the probability that among $k$ random candidates, the highest-ranked one is correct. Let $r_i$ denote the rank $i$ candidate and $\mathbf{1}[r_i]$ indicate correctness:
\begin{equation}\label{eq:best-of-k}
    \text{Bof}k := \frac{1}{{n \choose k}} \sum_{i=0}^{n-k}{n-i-1\choose k-1}\cdot \mathbf{1}[r_i \text{ is correct}].
\end{equation}

\paragraph{Throughput.} We measure the programs processed per second (PPS). Strong verifiers achieve high accuracy but low PPS; weak verifiers achieve high PPS but lower accuracy.

\paragraph{Verifiers.} The strongest verifier $\strongestVerifier{}$ (full test suite) runs all test cases and admits no false positives as a candidate that passes is correct. Weak verifiers $\weakVerifier{N}$ (first-$N$-tests) check only a subset of tests; they are fast but admit false positives. An outcome reward model $\ormVerifier{}$ (neural ranker) scores candidates without execution, offering constant-time inference regardless of test suite size.

\subsection{Staged Verification}\label{sec:staged-verification}

A program that fails \emph{any} test is incorrect; a program that passes \emph{all} tests is correct. This asymmetry motivates staged verification: use a cheap filter to eliminate obviously incorrect candidates before ranking survivors with an ORM. Removing candidates that can never be correct avoids wasted compute on ORM ranking and reduces possible spurious correlations, so with a sufficiently large ORM we would expect both accuracy and throughput improvements. 

We use deliberately simple filters—compilation, linting, execution on the first $N$ test cases—to isolate the interaction between filtering and neural ranking from the complexity of filter design. The pipeline is shown in \autoref{fig:pipeline}.

\subsection{ORMs as Verifiers}\label{sec:training-orm}
Let $\ormVerifier$ be the verifier that uses the ORM with parameters $\theta$ to predict if $c$ is correct. We denote its output as:
\begin{equation}\label{eq:orm}
    \ormVerifier(c \mid x) = P(c \text{ passes all tests} \mid x, c).
\end{equation}

We use a language model backbone with a scalar prediction head that outputs scores in $[0,1]$. Training uses pairs of candidates $(\candidate{w}, \candidate{\ell})$ where $\candidate{w}$ is correct and $\candidate{\ell}$ is incorrect. The Bradley-Terry objective~\citep{bradley_rank_1952,stiennon_learning_2020} maximizes the margin between correct and incorrect candidates. Formally, this objective is: 
\begin{equation}
    \label{eq:orm-objective}
    \mathcal{L}_{\text{RM}} = - \mathbb{E}_{(x,\candidate{w},\candidate{\ell}) \sim\mathcal{D}} [\log \sigma(\orm(\candidate{w} |x) - \orm(\candidate{\ell} |x))].
\end{equation}
At inference, the ORM scores each candidate in a single forward pass. Full training details are in \autoref{app:train-data}.

\paragraph{Why not process reward models?} Process reward models (PRMs) score individual reasoning steps rather than final outcomes~\citep{lightman_lets_2023,uesato_solving_2022}. We exclude PRMs for two reasons. First, step-level annotations are expensive. PRM800K required extensive human labeling~\citep{lightman_lets_2023}, and automated alternatives remain an active research area~\citep{wang_math-shepherd_2024}. Second, PRMs require scoring each step, resulting in multiple forward passes per candidate, directly undermining the throughput advantage that makes learned verifiers attractive. ORMs require only pass/fail labels and a single pass, making them practical for high-throughput verification.

%% file: sections/research_questions.tex
\section{Experimental Design}\label{sec:setup}

We train ORMs on CodeContests and GSM8K, then measure their accuracy and throughput across four benchmarks. In \autoref{sec:exp-setup} we describe the data, training, and evaluation protocol. In \autoref{sec:methods} we detail the ranking strategies and weak filters under comparison.

\subsection{Setup}\label{sec:exp-setup}

\paragraph{Data.} We train ORMs on CodeContests-Python~\citep{li_competition-level_2022} and GSM8K~\citep{cobbe_training_2021} in PAL format~\citep{gao_pal_2022}. Training candidates are generated with Qwen 2.5 Coder Instruct 7B~\citep{hui_qwen25-coder_2024} using VLLM~\citep{kwon_efficient_2023}, sampling $\trainNumSampled{}$ times per problem with temperature $T{=}1.0$ and $\text{top}_p{=}0.95$~\citep{holtzman_curious_2020}. We randomly sample up to 6 pairs per problem from 4.7K CodeContests and 7.3K GSM8K problems. For distribution shift evaluation, we use HumanEval~\citep{chen_evaluating_2021} and MBPP~\citep{austin_program_2021} via EvalPlus~\citep{liu_is_2023,liu_evaluating_2024}. All code is formatted using the Black formatter for standardization and to handle whitespace requirements.

To determine correctness, we follow existing works~\citep{cassano_multipl-e_2022,orlanski_measuring_2023,jain_livecodebench_2024} in using the subprocess\footnote{\url{https://docs.python.org/3/library/subprocess.html}} library to execute the programs with the test cases. For GSM8K, we check that the outputted scalar is correct using an assertion. In the case of CodeContests, we check that the program prints the correct output to STDOUT and stop execution on the first error. We detail our training data in \autoref{table:train-data}. We filter out any programs with syntax errors as these could be caused by the model hitting the token limit rather than a poor generation.

\paragraph{Training.} An ORM does not necessarily need to be a language model, but given the strong overall results of prior work in using a language model as an ORM~\citep{cobbe_training_2021}, we choose to use these. We train with the Bradley-Terry objective (\autoref{eq:orm-objective}), which models $P(\candidate{w} \succ \candidate{\ell})$ for correct candidate $\candidate{w}$ and incorrect candidate $\candidate{\ell}$. Intuitively, this objective pushes $\orm(\candidate{w} |x)$ higher than $\orm(\candidate{\ell} |x)$ by maximizing the margin between correct and incorrect candidates. At inference, the ORM scores any new candidate based on how likely it is to be preferred over incorrect solutions.
We train using Adam~\citep{kingma_adam_2017}, cosine scheduling (peak $5\text{e}{-}6$, 10\% warmup), batch size 64, and max sequence length 2048. We train for 2 epochs, use the final checkpoint, and average results over six random seeds. Both generators and ORM backbones are from the Qwen family. While this ensures compatibility, it may favor the ORM through distribution matching; we partially address this via distribution-shift evaluation on HumanEval and MBPP. Full hyperparameters are in \autoref{app:train-data}.

\paragraph{Evaluation.} CodeContests and GSM8K programs execute via subprocess with a 30s timeout~\citep{cassano_multipl-e_2022,orlanski_measuring_2023}. HumanEval and MBPP use EvalPlus~\citep{liu_is_2023,liu_evaluating_2024}. All verifiers are run across 5 independent trials to mitigate runtime noise. We measure preprocessing, execution, and postprocessing time, excluding disk I/O. $\ormVerifier{}$ experiments use a single 48GB A6000 GPU with 16 cores and 64GB memory. Execution-only setups use 64 cores and 256GB for CodeContests. For the other datasets, we use \numCoresUsed{} cores and 128GB memory. These configurations are roughly cost-equivalent according to the current prices per hour on AWS. We deduplicate candidates via exact string match after Black formatting, then weight outcomes by duplicate count for fair $\bestofk{k}$ measurement.

\paragraph{ORM Inference.} We use dynamic batching akin to sequence packing~\citep{JMLR:v21:20-074}: rather than fixed batch sizes bounded by outlier-length programs, we construct a batch by continuously adding candidates to the batch until the net tokens is over a threshold size. This establishes a fairer comparison to CPU methods as we are properly utilizing GPU memory and FLOPs. For the 500M ORM we set the max tokens per batch to 500,000 while for the 1.5B ORM we set the max tokens to 200,000.

\subsection{Verification Methods}\label{sec:methods}

We evaluate the following three ranking strategies and three filtering methods that can precede ranking:
\begin{enumerate}[noitemsep]
    \item \textbf{Majority Voting:} Select the most common candidate after normalization with Black. Standard method for ranking sequences generated~\citep{ehrlich_codemonkeys_2025}.
    \item $\mathbf{\ormVerifier{}}$\textbf{:} Score each candidate with the ORM and then select the highest-scoring candidate.
    \item $\mathbf{\strongestVerifier{}}$\textbf{:} Select the candidate passing the most tests.
\end{enumerate}

Next, for experiments involving staged verification, we use the following heuristics as weak verifiers: 
\begin{enumerate}[noitemsep]
    \item $\mathbf{\weakestVerifier{}}$\textbf{:} Remove candidates failing the syntax checker.
    \item $\mathbf{\lintVerifier{}}$\textbf{:} Remove candidates with PyLint errors.
    \item $\mathbf{\weakVerifier{N}}$\textbf{:} Remove candidates failing the first $N$ test cases within a timeout of 3 seconds. 
\end{enumerate}

We denote staged verification as $\pruneOrmVerifier{x}$ where $x$ is the filter. For example, $\tenTestVerifier{}$ filters with the first 10 tests, then ranks survivors with the ORM. For majority voting baselines, we rank by test cases passed with duplicate count as tiebreaker~\citep{ehrlich_codemonkeys_2025,li_competition-level_2022}.

%% file: sections/results.tex
\section{Results}
\input{tables/results_table.tex}

We first characterize when ORM-only ranking occupies favorable frontier points (\autoref{sec:orms}), then show how test-based filtering shifts the frontier further outward (\autoref{sec:staging}). \autoref{sec:mechanism} investigates why staging helps: ORMs systematically overrank candidates with subtle bugs that tests catch. \autoref{sec:test-selection} confirms that filtering effectiveness is robust to test choice, and \autoref{sec:practical} translates these findings into cost-accuracy trade-offs.

\autoref{fig:overview} shows that staged verification shifts the Pareto frontier outward across generator sizes. ORMs occupy favorable frontier points (\ormFasterPct{} faster than $\strongestVerifier{}$, \ormNaiveAccPct{} more accurate than naive baselines) across datasets (\autoref{tab:overall-results}) and generator sizes (\autoref{tab:codecontests-by-generator-results}). Staging improves on ORM-only points because lightweight filters remove incorrect candidates that ORMs systematically overrank. This benefit does not depend on specific test selection: random subsets achieve comparable false positive rates with low variance.

\input{sections/results_sections/orms.tex}

\input{sections/results_sections/pruning.tex}
\input{sections/results_sections/true_ranking.tex}
\input{sections/results_sections/test_randomness.tex}
\input{sections/results_sections/practical_implications.tex}

%% file: tables/results_table.tex
\begin{table*}[t]
    \centering

    \caption{Overall results for the different pruning with a weak verifier then ranking methods. If Filter is ``---'' that means no pruning is done. {\color{textgreen}{Green backgrounds}} is higher performance while {\color{textred}{Red backgrounds}} is lower performance with respect to the entire column. $\strongestVerifier{}$ is the case where \emph{all} test cases are run. Rows with Ranker set to ``MV'' are verifier-only baselines that select using majority voting without ORM inference. ``Syntax'' and ``Lint'' remove any programs with the respective errors. ``N Test'' prunes out any programs that do not pass the first $N$ test cases. The evaluation dataset is generated with Qwen 2.5 Coder 7B Instruct using $T=1.0$, $n=128$, $top_p=0.95$, and 1024 tokens. The results are averaged over six random seeds.} \label{tab:overall-results}
    \resizebox{0.8\textwidth}{!}{%
    \begin{tabular}{ll|cc|cc|cc|cc}
\toprule
 &  & \multicolumn{2}{c|}{CodeContests} & \multicolumn{2}{c|}{GSM8K} & \multicolumn{2}{c|}{HumanEval} & \multicolumn{2}{c}{MBPP} \\ Filter & Ranker & Bof64 & PPS & Bof64 & PPS & Bof64 & PPS & Bof64 & PPS \\
\midrule
\multirow[c]{3}{*}{---} & MV & \cellcolor[rgb]{0.722,0.510,0.510} 2.55 & \cellcolor[rgb]{0.510,0.722,0.569} 64777.78 & \cellcolor[rgb]{0.722,0.510,0.510} 83.28 & \cellcolor[rgb]{0.510,0.722,0.569} 39596.93 & \cellcolor[rgb]{1.000,1.000,1.000} 84.21 & \cellcolor[rgb]{0.510,0.722,0.569} 48077.73 & \cellcolor[rgb]{0.840,0.719,0.719} 66.36 & \cellcolor[rgb]{0.510,0.722,0.569} 52179.55 \\
 & 500M & \cellcolor[rgb]{0.861,0.754,0.754} 5.33 & \cellcolor[rgb]{0.883,0.793,0.793} 52.49 & \cellcolor[rgb]{0.974,0.954,0.954} 84.13 & \cellcolor[rgb]{1.000,1.000,1.000} 264.94 & \cellcolor[rgb]{0.724,0.515,0.515} 79.31 & \cellcolor[rgb]{0.995,0.997,0.996} 144.84 & \cellcolor[rgb]{0.794,0.637,0.637} 66.04 & \cellcolor[rgb]{0.929,0.959,0.937} 303.14 \\
 & 1.5B & \cellcolor[rgb]{0.973,0.953,0.953} 6.34 & \cellcolor[rgb]{0.722,0.510,0.510} 22.53 & \cellcolor[rgb]{0.755,0.861,0.784} 87.05 & \cellcolor[rgb]{0.809,0.664,0.664} 113.79 & \cellcolor[rgb]{0.837,0.713,0.713} 81.37 & \cellcolor[rgb]{0.807,0.659,0.659} 64.14 & \cellcolor[rgb]{0.976,0.958,0.958} 69.70 & \cellcolor[rgb]{0.834,0.707,0.707} 133.17 \\
\midrule
\multirow[c]{3}{*}{Syntax} & MV & \cellcolor[rgb]{0.722,0.510,0.510} 2.55 & \cellcolor[rgb]{0.732,0.848,0.764} 6122.37 & \cellcolor[rgb]{0.772,0.599,0.599} 83.32 & \cellcolor[rgb]{0.719,0.840,0.753} 6256.05 & \cellcolor[rgb]{0.999,0.999,0.999} 84.22 & \cellcolor[rgb]{0.723,0.843,0.756} 6774.73 & \cellcolor[rgb]{0.865,0.763,0.763} 66.63 & \cellcolor[rgb]{0.730,0.847,0.763} 5679.06 \\
 & 500M & \cellcolor[rgb]{0.912,0.844,0.844} 5.79 & \cellcolor[rgb]{0.956,0.923,0.923} 60.87 & \cellcolor[rgb]{0.960,0.977,0.965} 84.75 & \cellcolor[rgb]{0.951,0.972,0.957} 309.44 & \cellcolor[rgb]{0.739,0.540,0.540} 79.57 & \cellcolor[rgb]{0.995,0.997,0.996} 145.15 & \cellcolor[rgb]{0.722,0.510,0.510} 65.54 & \cellcolor[rgb]{0.880,0.932,0.895} 353.34 \\
 & 1.5B & \cellcolor[rgb]{1.000,1.000,1.000} 6.58 & \cellcolor[rgb]{0.735,0.533,0.533} 25.09 & \cellcolor[rgb]{0.510,0.722,0.569} 88.09 & \cellcolor[rgb]{0.861,0.755,0.755} 127.63 & \cellcolor[rgb]{0.894,0.813,0.813} 82.37 & \cellcolor[rgb]{0.833,0.705,0.705} 66.88 & \cellcolor[rgb]{0.978,0.987,0.981} 70.92 & \cellcolor[rgb]{0.873,0.777,0.777} 149.13 \\
\midrule
\multirow[c]{3}{*}{Lint} & MV & \cellcolor[rgb]{0.728,0.520,0.520} 2.67 & \cellcolor[rgb]{0.754,0.860,0.784} 302.09 & \cellcolor[rgb]{0.861,0.755,0.755} 83.39 & \cellcolor[rgb]{0.755,0.861,0.784} 488.17 & \cellcolor[rgb]{0.998,0.999,0.998} 84.23 & \cellcolor[rgb]{0.759,0.863,0.788} 633.15 & \cellcolor[rgb]{0.895,0.815,0.815} 67.44 & \cellcolor[rgb]{0.754,0.860,0.784} 613.41 \\
 & 500M & \cellcolor[rgb]{0.861,0.756,0.756} 5.34 & \cellcolor[rgb]{0.848,0.733,0.733} 47.56 & \cellcolor[rgb]{1.000,1.000,1.000} 84.30 & \cellcolor[rgb]{0.890,0.807,0.807} 156.85 & \cellcolor[rgb]{0.722,0.510,0.510} 79.26 & \cellcolor[rgb]{0.928,0.873,0.873} 101.10 & \cellcolor[rgb]{0.842,0.722,0.722} 66.37 & \cellcolor[rgb]{0.901,0.826,0.826} 166.69 \\
 & 1.5B & \cellcolor[rgb]{0.999,1.000,1.000} 6.60 & \cellcolor[rgb]{0.722,0.511,0.511} 22.64 & \cellcolor[rgb]{0.755,0.861,0.784} 87.05 & \cellcolor[rgb]{0.722,0.510,0.510} 90.24 & \cellcolor[rgb]{0.886,0.799,0.799} 82.23 & \cellcolor[rgb]{0.722,0.510,0.510} 55.19 & \cellcolor[rgb]{1.000,1.000,1.000} 70.36 & \cellcolor[rgb]{0.722,0.510,0.510} 100.38 \\
\midrule
\multirow[c]{3}{*}{1 Test} & MV & \cellcolor[rgb]{0.770,0.869,0.797} 15.72 & \cellcolor[rgb]{0.755,0.861,0.784} 131.73 & --- & --- & \cellcolor[rgb]{0.719,0.841,0.753} 87.62 & \cellcolor[rgb]{0.753,0.860,0.783} 1020.54 & \cellcolor[rgb]{0.774,0.872,0.801} 76.07 & \cellcolor[rgb]{0.753,0.860,0.782} 930.69 \\
 & 500M & \cellcolor[rgb]{0.769,0.869,0.797} 15.75 & \cellcolor[rgb]{0.985,0.991,0.987} 69.50 & --- & --- & \cellcolor[rgb]{0.981,0.966,0.966} 83.88 & \cellcolor[rgb]{0.981,0.966,0.966} 125.86 & \cellcolor[rgb]{0.810,0.892,0.833} 75.17 & \cellcolor[rgb]{1.000,1.000,1.000} 229.08 \\
 & 1.5B & \cellcolor[rgb]{0.677,0.817,0.716} 16.86 & \cellcolor[rgb]{0.934,0.884,0.884} 58.35 & --- & --- & \cellcolor[rgb]{0.833,0.905,0.853} 85.98 & \cellcolor[rgb]{0.799,0.646,0.646} 63.32 & \cellcolor[rgb]{0.743,0.854,0.774} 77.04 & \cellcolor[rgb]{0.831,0.703,0.703} 132.52 \\
\midrule
\multirow[c]{3}{*}{10 Tests} & MV & \cellcolor[rgb]{0.641,0.796,0.684} 17.12 & \cellcolor[rgb]{0.777,0.873,0.804} 120.93 & --- & --- & \cellcolor[rgb]{0.510,0.722,0.569} 92.46 & \cellcolor[rgb]{0.755,0.861,0.784} 650.77 & \cellcolor[rgb]{0.522,0.729,0.580} 86.06 & \cellcolor[rgb]{0.888,0.936,0.902} 345.21 \\
 & 500M & \cellcolor[rgb]{0.582,0.763,0.633} 17.54 & \cellcolor[rgb]{0.970,0.983,0.974} 73.22 & --- & --- & \cellcolor[rgb]{0.616,0.782,0.662} 90.01 & \cellcolor[rgb]{1.000,1.000,1.000} 134.92 & \cellcolor[rgb]{0.544,0.741,0.598} 85.19 & \cellcolor[rgb]{0.905,0.833,0.833} 169.30 \\
 & 1.5B & \cellcolor[rgb]{0.510,0.722,0.569} 18.06 & \cellcolor[rgb]{1.000,1.000,1.000} 65.80 & --- & --- & \cellcolor[rgb]{0.614,0.781,0.660} 90.05 & \cellcolor[rgb]{0.867,0.766,0.766} 72.83 & \cellcolor[rgb]{0.510,0.722,0.569} 86.57 & \cellcolor[rgb]{0.793,0.636,0.636} 121.29 \\
\midrule
\multicolumn{2}{c|}{$\strongestVerifier{}$} & \cellcolor[rgb]{0.510,0.722,0.569} 20.69 & \cellcolor[rgb]{0.722,0.510,0.510} 2.95 & \cellcolor[rgb]{0.510,0.722,0.569} 97.96 & \cellcolor[rgb]{0.722,0.510,0.510} 88.21 & \cellcolor[rgb]{0.510,0.722,0.569} 95.97 & \cellcolor[rgb]{0.722,0.510,0.510} 22.19 & \cellcolor[rgb]{0.510,0.722,0.569} 90.04 & \cellcolor[rgb]{0.722,0.510,0.510} 15.01 \\
\bottomrule
\end{tabular}

    }
\end{table*}

%% file: sections/results_sections/orms.tex
\newcommand{\smallORMFasterLargePct}{1.44$\times$}
\newcommand{\smallORMWorseLargePct}{4.99\%}
\subsection{ORMs Occupy Favorable Frontier Points}\label{sec:orms}

\autoref{tab:overall-results} summarizes accuracy (\bestofk{64}) and throughput (PPS) across verification methods. Results are averaged over six random seeds with 95\% confidence intervals; median CI width is 1.48 percentage points.

When full-suite execution is the bottleneck, $\ormVerifier{}$ achieves \ormFasterPct{} higher throughput than $\strongestVerifier{}$ while remaining more accurate than naive baselines (\autoref{tab:overall-results}), making ORM-only ranking a practical frontier choice. When even partial execution is inexpensive, however, running a small number of tests provides reliable signal at low cost, and ORM-only points become less competitive. Smaller ORMs sharpen this trade-off: the 500M variant is \smallORMFasterLargePct{} faster than the 1.5B model with only a \smallORMWorseLargePct{} accuracy loss. \textbf{ORM-only ranking is most compelling when execution is expensive; when even partial execution is cheap, ORM-only points become less competitive on the frontier.}

%% file: sections/results_sections/pruning.tex
\newcommand{\simpleFilterFiveHundoPPS}{10.42$\times$}
\newcommand{\simpleFilterOneFivePPS}{4.24$\times$}
\newcommand{\fiverHundoBof}{45.15\%}
\newcommand{\oneFiveBof}{38.09\%}
\newcommand{\testVerifierImproveBof}{69.23\%}
\newcommand{\testVerifierImprovePPS}{12.41$\times$}
\newcommand{\testLargeVerifierOrmSpeed}{57.64\%}
\newcommand{\tenTestImproveBof}{69.92\%}
\newcommand{\tenTestWorseStrongBof}{7.58\%}
\newcommand{\tenTestFaster}{10.22$\times$}
\newcommand{\wvGenBof}{7.64\%}
\newcommand{\wvGenPPS}{8.65$\times$}
\newcommand{\wvSmallTempPPS}{146.40}
\newcommand{\wvSmallTempPPSSTD}{68.97}
\newcommand{\wvLargeTempPPS}{89.28}
\newcommand{\wvLargeTempPPSSTD}{33.49}
\newcommand{\wvTempPPSImprove}{13.44$\times$}
\newcommand{\wvTempBofWorse}{8.80}
\newcommand{\syntaxPctPruned}{1.30\%}
\newcommand{\lintPctPruned}{2.62\%}
\newcommand{\wvTenRemovalMin}{90\%}
\newcommand{\wvTenRemovalMax}{97\%}
\newcommand{\tenTestLargeOrmAcc}{18.06\%}
\newcommand{\tenTestLargeOrmCILow}{17.67}
\newcommand{\tenTestLargeOrmCIHigh}{18.45}
\newcommand{\tenTestAccGain}{12}
\newcommand{\accGainCIMultiple}{15}
\newcommand{\ppsCIPctCC}{1\%}
\newcommand{\ppsCIPctOther}{3--10\%}
\newcommand{\speedupCIMultiple}{10}

\subsection{Staged Verification Shifts the Frontier}\label{sec:staging}

\input{tables/codecontests_by_generator.tex}
\autoref{tab:codecontests-by-generator-results} shows results across generators. Syntax and lint filtering remove only \syntaxPctPruned{} and \lintPctPruned{} of candidates on average as shown in \autoref{tab:pruning_tokens}. The filtering overhead exceeds inference savings, decreasing throughput by \simpleFilterFiveHundoPPS{} for the 500M ORM and \simpleFilterOneFivePPS{} for the 1.5B. Test-based filtering changes the picture: on CodeContests, $\weakVerifier{10}$ removes \wvTenRemovalMin{} to \wvTenRemovalMax{} of candidates, shifting the cost balance so that $\nTestPruneVerifier{}$ delivers \testVerifierImprovePPS{} higher throughput than $\strongestVerifier{}$. \textbf{The best configuration, $\tenTestVerifier{}$ with the 1.5B ORM, is \tenTestFaster{} faster than $\strongestVerifier{}$ while losing only \tenTestWorseStrongBof{} accuracy.}

Staged verification is not purely a speed optimization, as accuracy \emph{improves} alongside throughput: the 500M ORM sees a \fiverHundoBof{} rise in $\bestofk{64}$ and the 1.5B a \oneFiveBof{} rise over their unfiltered counterparts because \textbf{test filtering removes candidates the ORM would have misranked}. On CodeContests, the 1.5B ORM with $\tenTestVerifier{}$ achieves \tenTestLargeOrmAcc{} with 95\% CI of [\tenTestLargeOrmCILow, \tenTestLargeOrmCIHigh]. These accuracy gains of \tenTestAccGain{}+ points over ORM-only exceed CI widths by roughly \accGainCIMultiple{}$\times$.

These patterns hold across generator sizes and sampling temperatures. With the 14B generator, $\tenTestVerifier{}$ yields \wvGenPPS{} speedup at \wvGenBof{} accuracy loss. Temperature ablations show average throughput of \wvSmallTempPPS{} $\pm$\wvSmallTempPPSSTD{} and \wvLargeTempPPS{} $\pm$\wvLargeTempPPSSTD{} PPS for the 500M and 1.5B ORMs, which is \wvTempPPSImprove{} faster than $\strongestVerifier{}$ while accuracy degrades by only \wvTempBofWorse{} points. Throughput CIs are tight for CodeContests at less than \ppsCIPctCC{} of mean and moderate for HumanEval/MBPP at \ppsCIPctOther{}. The latter use EvalPlus, which executes code via \texttt{eval} rather than subprocess, introducing additional runtime variance. Reported speedups exceed CI widths by \speedupCIMultiple{}$\times$ or more. ORMs remain most useful when partial execution has high fixed costs like environment setup or container launch, or when filtering leaves a large ambiguous survivor set. \textbf{We find that staged verification shifts the Pareto frontier outward, with $\tenTestVerifier{}$ achieving near-$\strongestVerifier{}$ accuracy at an order of magnitude higher throughput.}

%% file: tables/codecontests_by_generator.tex
\begin{table*}[t]
    \centering

    \caption{Results for the different pruning then ranking on only CodeContests as we scale the size of the generator model. If Filter is ``---'' that means no pruning is done. {\color{textgreen}{Green backgrounds}} is higher performance while {\color{textred}{Red backgrounds}} is lower performance with respect to the entire column. $\strongestVerifier{}$ is the case where \emph{all} test cases are run. Rows with Ranker set to ``MV'' are verifier-only baselines that select using majority voting without ORM inference. ``Syntax'' and ``Lint'' remove any programs with the respective errors. ``N Test'' prunes out any programs that do not pass the first $N$ test cases. The evaluation dataset is generated with Qwen 2.5 Coder Instruct models using $T=1.0$, $n=128$, $top_p=0.95$, and 1024 tokens. The results are averaged over six random seeds.}\label{tab:codecontests-by-generator-results}
    \resizebox{0.9\textwidth}{!}{%
    \begin{tabular}{ll|cc|cc|cc|cc|cc}
\toprule
 &  & \multicolumn{2}{c|}{500M} & \multicolumn{2}{c|}{1.5B} & \multicolumn{2}{c|}{3B} & \multicolumn{2}{c|}{7B} & \multicolumn{2}{c}{14B} \\ Filter & Ranker & Bof64 & PPS & Bof64 & PPS & Bof64 & PPS & Bof64 & PPS & Bof64 & PPS \\
\midrule
\multirow[c]{3}{*}{---} & MV & \cellcolor[rgb]{0.722,0.510,0.510} 0.00 & \cellcolor[rgb]{0.510,0.722,0.569} 65679.19 & \cellcolor[rgb]{0.722,0.510,0.510} 0.09 & \cellcolor[rgb]{0.510,0.722,0.569} 66369.90 & \cellcolor[rgb]{0.722,0.510,0.510} 1.87 & \cellcolor[rgb]{0.510,0.722,0.569} 64541.54 & \cellcolor[rgb]{0.722,0.510,0.510} 2.55 & \cellcolor[rgb]{0.510,0.722,0.569} 64777.78 & \cellcolor[rgb]{0.722,0.510,0.510} 2.21 & \cellcolor[rgb]{0.510,0.722,0.569} 62487.39 \\
 & 500M & \cellcolor[rgb]{0.866,0.763,0.763} 0.47 & \cellcolor[rgb]{0.903,0.829,0.829} 68.39 & \cellcolor[rgb]{0.867,0.766,0.766} 1.64 & \cellcolor[rgb]{0.924,0.866,0.866} 67.79 & \cellcolor[rgb]{0.872,0.775,0.775} 4.52 & \cellcolor[rgb]{0.977,0.960,0.960} 64.02 & \cellcolor[rgb]{0.861,0.754,0.754} 5.33 & \cellcolor[rgb]{0.883,0.793,0.793} 52.49 & \cellcolor[rgb]{0.942,0.897,0.897} 10.99 & \cellcolor[rgb]{0.975,0.957,0.957} 59.83 \\
 & 1.5B & \cellcolor[rgb]{1.000,1.000,1.000} 0.61 & \cellcolor[rgb]{0.722,0.510,0.510} 26.42 & \cellcolor[rgb]{1.000,1.000,1.000} 2.62 & \cellcolor[rgb]{0.722,0.510,0.510} 26.63 & \cellcolor[rgb]{0.996,0.998,0.997} 5.76 & \cellcolor[rgb]{0.729,0.523,0.523} 25.14 & \cellcolor[rgb]{0.973,0.953,0.953} 6.34 & \cellcolor[rgb]{0.722,0.510,0.510} 22.53 & \cellcolor[rgb]{1.000,1.000,1.000} 12.86 & \cellcolor[rgb]{0.739,0.540,0.540} 23.52 \\
\midrule
\multirow[c]{3}{*}{Syntax} & MV & \cellcolor[rgb]{0.722,0.510,0.510} 0.00 & \cellcolor[rgb]{0.734,0.849,0.766} 5642.49 & \cellcolor[rgb]{0.722,0.510,0.510} 0.09 & \cellcolor[rgb]{0.734,0.849,0.766} 5785.38 & \cellcolor[rgb]{0.722,0.510,0.510} 1.87 & \cellcolor[rgb]{0.736,0.850,0.768} 4972.95 & \cellcolor[rgb]{0.722,0.510,0.510} 2.55 & \cellcolor[rgb]{0.732,0.848,0.764} 6122.37 & \cellcolor[rgb]{0.790,0.629,0.629} 5.23 & \cellcolor[rgb]{0.730,0.847,0.763} 6432.92 \\
 & 500M & \cellcolor[rgb]{0.859,0.752,0.752} 0.46 & \cellcolor[rgb]{0.870,0.771,0.771} 64.53 & \cellcolor[rgb]{0.868,0.768,0.768} 1.65 & \cellcolor[rgb]{0.962,0.932,0.932} 70.63 & \cellcolor[rgb]{0.876,0.781,0.781} 4.55 & \cellcolor[rgb]{1.000,1.000,1.000} 65.86 & \cellcolor[rgb]{0.912,0.844,0.844} 5.79 & \cellcolor[rgb]{0.956,0.923,0.923} 60.87 & \cellcolor[rgb]{0.943,0.899,0.899} 11.02 & \cellcolor[rgb]{1.000,1.000,1.000} 62.31 \\
 & 1.5B & \cellcolor[rgb]{0.971,0.949,0.949} 0.58 & \cellcolor[rgb]{0.724,0.515,0.515} 27.16 & \cellcolor[rgb]{0.996,0.998,0.996} 2.67 & \cellcolor[rgb]{0.728,0.521,0.521} 28.25 & \cellcolor[rgb]{1.000,1.000,1.000} 5.68 & \cellcolor[rgb]{0.731,0.526,0.526} 25.56 & \cellcolor[rgb]{1.000,1.000,1.000} 6.58 & \cellcolor[rgb]{0.735,0.533,0.533} 25.09 & \cellcolor[rgb]{0.997,0.998,0.997} 13.03 & \cellcolor[rgb]{0.744,0.550,0.550} 24.61 \\
\midrule
\multirow[c]{3}{*}{Lint} & MV & \cellcolor[rgb]{0.722,0.510,0.510} 0.00 & \cellcolor[rgb]{0.753,0.860,0.783} 523.78 & \cellcolor[rgb]{0.722,0.511,0.511} 0.10 & \cellcolor[rgb]{0.753,0.860,0.783} 550.65 & \cellcolor[rgb]{0.724,0.515,0.515} 1.92 & \cellcolor[rgb]{0.753,0.860,0.783} 548.99 & \cellcolor[rgb]{0.728,0.520,0.520} 2.67 & \cellcolor[rgb]{0.754,0.860,0.784} 302.09 & \cellcolor[rgb]{0.790,0.630,0.630} 5.25 & \cellcolor[rgb]{0.754,0.860,0.783} 459.62 \\
 & 500M & \cellcolor[rgb]{0.866,0.763,0.763} 0.47 & \cellcolor[rgb]{0.857,0.748,0.748} 62.47 & \cellcolor[rgb]{0.857,0.748,0.748} 1.55 & \cellcolor[rgb]{0.844,0.726,0.726} 58.69 & \cellcolor[rgb]{0.855,0.745,0.745} 4.31 & \cellcolor[rgb]{0.845,0.727,0.727} 51.05 & \cellcolor[rgb]{0.861,0.756,0.756} 5.34 & \cellcolor[rgb]{0.848,0.733,0.733} 47.56 & \cellcolor[rgb]{0.849,0.735,0.735} 7.89 & \cellcolor[rgb]{0.840,0.719,0.719} 44.11 \\
 & 1.5B & \cellcolor[rgb]{0.986,0.992,0.988} 0.67 & \cellcolor[rgb]{0.740,0.542,0.542} 31.28 & \cellcolor[rgb]{1.000,1.000,1.000} 2.62 & \cellcolor[rgb]{0.724,0.514,0.514} 27.19 & \cellcolor[rgb]{0.977,0.959,0.959} 5.47 & \cellcolor[rgb]{0.722,0.510,0.510} 23.45 & \cellcolor[rgb]{0.999,1.000,1.000} 6.60 & \cellcolor[rgb]{0.722,0.511,0.511} 22.64 & \cellcolor[rgb]{0.877,0.783,0.783} 8.90 & \cellcolor[rgb]{0.722,0.510,0.510} 20.01 \\
\midrule
\multirow[c]{3}{*}{1 Test} & MV & \cellcolor[rgb]{0.821,0.899,0.843} 1.39 & \cellcolor[rgb]{0.755,0.861,0.784} 126.19 & \cellcolor[rgb]{0.812,0.893,0.835} 4.92 & \cellcolor[rgb]{0.755,0.861,0.784} 124.94 & \cellcolor[rgb]{0.822,0.899,0.844} 9.65 & \cellcolor[rgb]{0.755,0.861,0.784} 122.58 & \cellcolor[rgb]{0.770,0.869,0.797} 15.72 & \cellcolor[rgb]{0.755,0.861,0.784} 131.73 & \cellcolor[rgb]{0.800,0.886,0.824} 22.76 & \cellcolor[rgb]{0.755,0.861,0.784} 132.92 \\
 & 500M & \cellcolor[rgb]{0.739,0.852,0.771} 1.71 & \cellcolor[rgb]{1.000,1.000,1.000} 79.64 & \cellcolor[rgb]{0.740,0.852,0.771} 5.64 & \cellcolor[rgb]{0.997,0.999,0.998} 74.06 & \cellcolor[rgb]{0.765,0.867,0.793} 10.93 & \cellcolor[rgb]{0.996,0.998,0.997} 66.69 & \cellcolor[rgb]{0.769,0.869,0.797} 15.75 & \cellcolor[rgb]{0.985,0.991,0.987} 69.50 & \cellcolor[rgb]{0.766,0.867,0.794} 24.47 & \cellcolor[rgb]{0.985,0.992,0.987} 66.16 \\
 & 1.5B & \cellcolor[rgb]{0.578,0.760,0.628} 2.02 & \cellcolor[rgb]{0.954,0.918,0.918} 74.27 & \cellcolor[rgb]{0.757,0.862,0.786} 5.60 & \cellcolor[rgb]{0.918,0.856,0.856} 67.36 & \cellcolor[rgb]{0.725,0.844,0.758} 11.39 & \cellcolor[rgb]{0.904,0.830,0.830} 58.02 & \cellcolor[rgb]{0.677,0.817,0.716} 16.86 & \cellcolor[rgb]{0.934,0.884,0.884} 58.35 & \cellcolor[rgb]{0.713,0.837,0.748} 25.52 & \cellcolor[rgb]{0.902,0.828,0.828} 52.46 \\
\midrule
\multirow[c]{3}{*}{10 Tests} & MV & \cellcolor[rgb]{0.762,0.865,0.790} 1.65 & \cellcolor[rgb]{0.756,0.862,0.786} 125.65 & \cellcolor[rgb]{0.673,0.814,0.712} 5.73 & \cellcolor[rgb]{0.761,0.864,0.790} 122.34 & \cellcolor[rgb]{0.625,0.787,0.670} 12.16 & \cellcolor[rgb]{0.770,0.869,0.797} 116.15 & \cellcolor[rgb]{0.641,0.796,0.684} 17.12 & \cellcolor[rgb]{0.777,0.873,0.804} 120.93 & \cellcolor[rgb]{0.689,0.823,0.726} 25.83 & \cellcolor[rgb]{0.778,0.874,0.805} 120.53 \\
 & 500M & \cellcolor[rgb]{0.713,0.837,0.748} 1.76 & \cellcolor[rgb]{0.959,0.976,0.964} 87.47 & \cellcolor[rgb]{0.510,0.722,0.569} 5.95 & \cellcolor[rgb]{0.972,0.984,0.976} 79.18 & \cellcolor[rgb]{0.558,0.749,0.611} 12.67 & \cellcolor[rgb]{0.977,0.987,0.980} 70.78 & \cellcolor[rgb]{0.582,0.763,0.633} 17.54 & \cellcolor[rgb]{0.970,0.983,0.974} 73.22 & \cellcolor[rgb]{0.516,0.725,0.574} 28.02 & \cellcolor[rgb]{0.982,0.990,0.984} 67.14 \\
 & 1.5B & \cellcolor[rgb]{0.510,0.722,0.569} 2.15 & \cellcolor[rgb]{0.986,0.992,0.987} 82.33 & \cellcolor[rgb]{0.658,0.806,0.699} 5.75 & \cellcolor[rgb]{1.000,1.000,1.000} 73.53 & \cellcolor[rgb]{0.510,0.722,0.569} 13.04 & \cellcolor[rgb]{0.979,0.964,0.964} 64.18 & \cellcolor[rgb]{0.510,0.722,0.569} 18.06 & \cellcolor[rgb]{1.000,1.000,1.000} 65.80 & \cellcolor[rgb]{0.510,0.722,0.569} 28.10 & \cellcolor[rgb]{0.952,0.915,0.915} 57.46 \\
\midrule
\multicolumn{2}{c|}{$\strongestVerifier{}$} & \cellcolor[rgb]{0.510,0.722,0.569} 2.16 & \cellcolor[rgb]{0.722,0.510,0.510} 3.32 & \cellcolor[rgb]{0.510,0.722,0.569} 7.83 & \cellcolor[rgb]{0.722,0.510,0.510} 2.23 & \cellcolor[rgb]{0.510,0.722,0.569} 14.92 & \cellcolor[rgb]{0.722,0.510,0.510} 2.54 & \cellcolor[rgb]{0.510,0.722,0.569} 20.69 & \cellcolor[rgb]{0.722,0.510,0.510} 2.95 & \cellcolor[rgb]{0.510,0.722,0.569} 31.98 & \cellcolor[rgb]{0.722,0.510,0.510} 3.65 \\
\bottomrule
\end{tabular}

    }
\end{table*}

%% file: sections/results_sections/true_ranking.tex
\newcommand{\syntaxTestFiveLowest}{96.72}
\newcommand{\oneTestFiveLowest}{65.29}
\newcommand{\tenTestFiveLowest}{65.28}
\newcommand{\smallModelSyntaxSpread}{35.36}
\newcommand{\smallModelSingleSpread}{29.99}
\newcommand{\smallModelTenSpread}{30.30}
\newcommand{\largeModelSyntaxSpread}{33.74}
\newcommand{\largeModelSingleSpread}{28.17}
\newcommand{\largeModelTenSpread}{28.53}
\newcommand{\smallOneTestPPS}{157.23}
\newcommand{\smallTenTestPPS}{120.62}
\newcommand{\largeOneTestPPS}{89.58}
\newcommand{\largeTenTestPPS}{83.69}
\newcommand{\smallOneTestPPSSTD}{77.66}
\newcommand{\smallTenTestPPSSTD}{33.68}
\newcommand{\largeOneTestPPSSTD}{37.29}
\newcommand{\largeTenTestPPSSTD}{22.41}
\newcommand{\smallAccStdDrop}{23.23\%}
\newcommand{\largeAccStdDrop}{31.23\%}
\newcommand{\rankDifference}{31}
\newcommand{\topEnrichWhileLen}{46}
\newcommand{\topEnrichLargeNum}{40}
\newcommand{\topEnrichRemove}{22}

\subsection{ORMs Systematically Overrank Small Bugs}\label{sec:mechanism}
\begin{figure*}[h]
    \centering
    \includegraphics[width=1\textwidth]{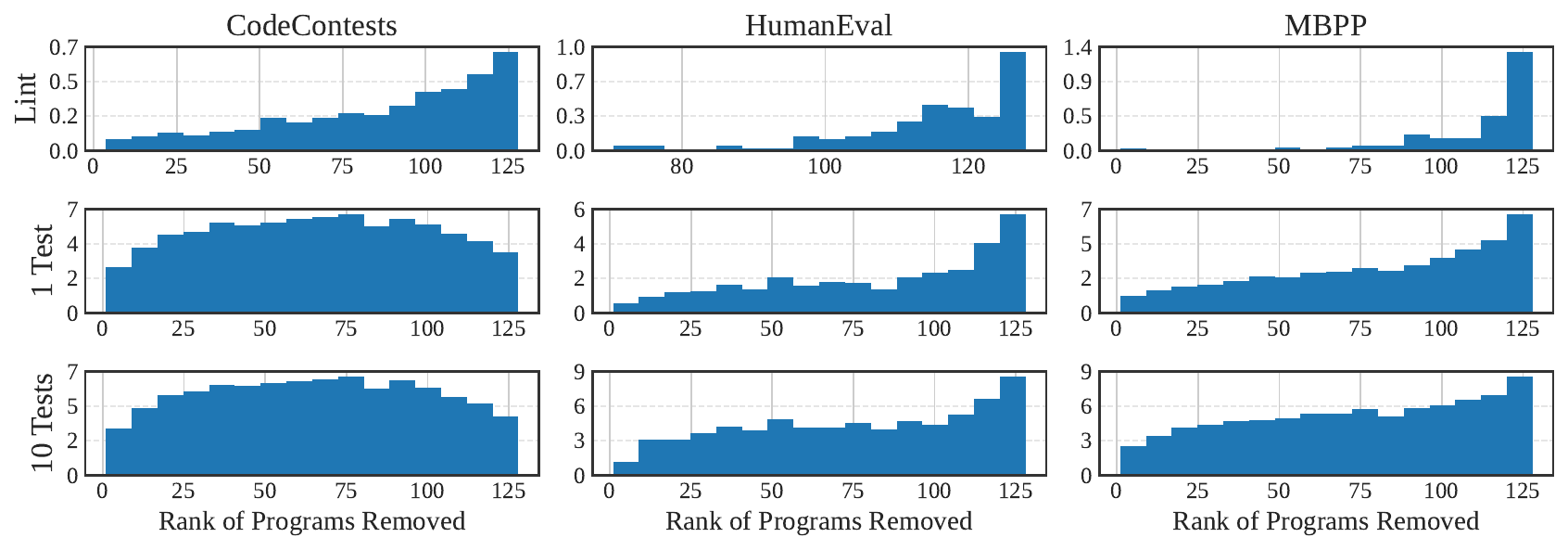}
    \caption{Distribution of filtered candidates by ORM rank. Rank 1 is the top-ranked candidate; rank 128 is the lowest. Rows are filters; columns are datasets. Results shown for the 1.5B ORM; the 500M ORM graph is in \autoref{fig:true-ranking-qc-inst-500m-t10-n128}.}
    \label{fig:true-ranking-qc-inst-7b-t10-n128}
\end{figure*}

We do not retrain the ORM after filtering; the only way to improve $\bestofk{k}$ is to remove incorrect candidates that the ORM ranked too highly. In \autoref{fig:true-ranking-qc-inst-7b-t10-n128} we show where the filtered candidates fall in the ORM ranking.

Syntax-filtered candidates cluster at the bottom (mean rank \syntaxTestFiveLowest{} out of 128, placing them in the lowest quartile). Removing candidates the ORM already dislikes changes little. Test-filtered candidates tell a different story: mean rank \oneTestFiveLowest{}, a shift of \rankDifference{} positions toward the top. These are incorrect candidates the ORM ranked in the middle or higher; removing them directly improves accuracy. Filtering also stabilizes results: with $\tenTestVerifier{}$, $\bestofk{64}$ standard deviation drops by \smallAccStdDrop{} (500M) and \largeAccStdDrop{} (1.5B).

\paragraph{What bugs do ORMs miss?}
We analyze candidates the ORM ranked in its top-16 but $\weakVerifier{1}$ filtered. Comparing pattern prevalence in these ``overrated'' candidates versus correct ones reveals which bugs ORMs fail to penalize.

\input{tables/bug_patterns_by_dataset.tex}

Three patterns are 20 to 46$\times$ more common in overrated candidates: \texttt{while len(x)} loops (\topEnrichWhileLen{}$\times$ enrichment), large integer arithmetic (\topEnrichLargeNum{}$\times$), and \texttt{.remove()} inside iteration (\topEnrichRemove{}$\times$). \autoref{tab:bug-patterns-by-dataset} confirms the mechanism: these patterns appear at similar rates in the full candidate pool and in the ORM's top-16 selections.

These bugs share a property: they preserve surface form. Float division (\texttt{/}) and integer division (\texttt{//}) differ by one character. \texttt{int(x/2)} and \texttt{x//2} look similar but behave differently for negative numbers. ORMs trained on correctness labels see similar token sequences and assign similar scores; a single test catches the runtime difference. \textbf{Test-based filtering catches bugs the ORM misses, particularly single-character errors that preserve surface form but fail at runtime. Filtered candidates sat \rankDifference{} ranks too high on average; removing them shifts the frontier.}

%% file: tables/bug_patterns_by_dataset.tex
\begin{table}[t]
\centering
\small
\caption{Bug patterns the ORM fails to penalize. Freq is overall pattern prevalence in problems with at least one passing solution; Top-16 is the percentage of solutions in the top-16 ORM-rated selections containing the pattern.}
\label{tab:bug-patterns-by-dataset}
\begin{tabular}{lrr}
\toprule
\multicolumn{3}{c}{\textit{CodeContests}} \\
\midrule
\textbf{Pattern} & \textbf{Freq} & \textbf{Top-16} \\
\midrule
\texttt{sorted()} w/o key & 4.8\% & 2.5\% \\
\texttt{/} instead of \texttt{//} & 2.9\% & 3.4\% \\
\texttt{.remove()} in loop & 1.3\% & 2.1\% \\
Large number ops & 1.1\% & 0.8\% \\
\texttt{while len()} & 0.7\% & 0.8\% \\
\midrule
\\
\multicolumn{3}{c}{\textit{HumanEval + MBPP}} \\
\midrule
\textbf{Pattern} & \textbf{Freq} & \textbf{Top-16} \\
\midrule
\texttt{/} instead of \texttt{//} & 3.8\% & 3.3\% \\
\texttt{sorted()} w/o key & 2.1\% & 3.7\% \\
\texttt{/2} in binary search & 1.5\% & 0.7\% \\
\texttt{(a+b)/2} midpoint & 0.6\% & 0.3\% \\
\bottomrule
\end{tabular}
\end{table}

%% file: sections/results_sections/test_randomness.tex
\subsection{Robustness to Test Selection}\label{sec:test-selection}

How sensitive is filtering effectiveness to which tests are selected? If there are large discrepancies between the discriminative power of test cases, then using staged verification for code generation is not generalizable. We investigate the false-positive rate (FPR) of test cases on CodeContests to determine if our previously discussed results are an artifact of lucky test selection. To do this, we profile the test execution time of generated solutions for CodeContests with the 1.5B and 3B Qwen 2.5 Coder Instruct generated solutions.\footnote{The 7B and 14B generated solutions continuously caused our 64 core/256GB to hang indefinitely with the additional profiling.} We run each test case five times with the solution and measure its mean wall time to run in isolation.

\begin{figure}[t]
\centering
\includegraphics[width=\columnwidth]{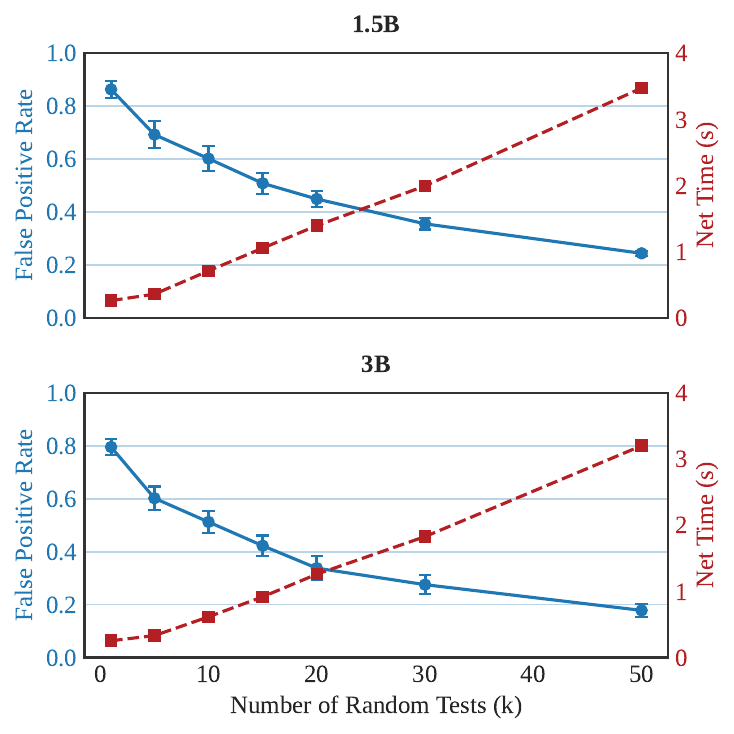}
\caption{False positive rate and cumulative time for random test subsets (top: 1.5B, bottom: 3B). Error bars show within-problem standard deviation across subset selections.}
\label{fig:test-randomness}
\end{figure}

In \autoref{fig:test-randomness} we show that test selection has minimal impact. At $k=1$, mean FPR is 86.3\% (1.5B) and 79.6\% (3B), with low variance across test choices (error bars). At $k=10$, mean FPR drops to 60.1\% in 0.71\,s (1.5B) and 51.2\% in 0.62\,s (3B). Incorrect LLM-generated code tends to fail broadly rather than on specific edge cases---any single test is roughly as effective as any other. We conclude that the filtering benefit is not be an artifact of test ordering, but a generalizable property of LLM bug distributions.

Our error bars show within-problem standard deviations across five random subsets. For FPR, this is 3.3 and 3.1 percentage points (1.5B/3B) at $k=1$, and 4.8 and 4.0 points at $k=10$, corresponding to 95\% confidence intervals of roughly $\pm$4--6 points. This suggests more efficient test-suite verification techniques than running a full test suite on all candidates---e.g. filtering with a small random subset of tests and running the full suite on the candidates that survive the filter---which we leave as a promising future direction. \textbf{We find staged verification to not require careful test curation, as random test subsets achieve comparable false positive rates with low variance.}

%% file: sections/results_sections/practical_implications.tex
\newcommand{\tenTestMVCost}{\$0.69}
\newcommand{\tenTestMVCostCILow}{\$0.68}
\newcommand{\tenTestMVCostCIHigh}{\$0.70}
\newcommand{\tenTestOrmCostMin}{\$1.3}
\newcommand{\tenTestOrmCostMax}{\$2.4}
\newcommand{\strongestVerifierCost}{\$28}
\newcommand{\stagedCostRatioMin}{12}
\newcommand{\stagedCostRatioMax}{22}
\newcommand{\oneTestMVCost}{\$0.63}
\newcommand{\oneTestMVCostCILow}{\$0.63}
\newcommand{\oneTestMVCostCIHigh}{\$0.64}
\newcommand{\oneTestOrmCostMin}{\$1.4}
\newcommand{\oneTestOrmCostMax}{\$2.8}
\newcommand{\oneTestTenTestCostDiff}{\$0.06}
\newcommand{\strongestVerifierAcc}{20.69\%}
\newcommand{\strongestVerifierAccCILow}{20.28}
\newcommand{\strongestVerifierAccCIHigh}{21.11}
\newcommand{\oneTestMVAcc}{15.72\%}
\newcommand{\oneTestMVAccCILow}{15.44}
\newcommand{\oneTestMVAccCIHigh}{16.00}
\newcommand{\tenTestMVAcc}{17.12\%}
\newcommand{\tenTestMVAccCILow}{16.80}
\newcommand{\tenTestMVAccCIHigh}{17.44}
\newcommand{\oneTestTenTestAccGain}{1.40\%}
\newcommand{\oneTestPPSMin}{122.6}
\newcommand{\oneTestPPSMax}{132.9}
\newcommand{\tenTestPPSMin}{116.2}
\newcommand{\tenTestPPSMax}{125.7}

\subsection{Practical Implications}\label{sec:practical}

Our findings enable cost-accuracy Pareto frontiers based on cloud provider pricing (\autoref{tab:cloud-costs}); in \autoref{fig:cost-tradeoff-by-generator} we show this frontier for CodeContests, where $\tenTestVerifier{}$ is Pareto-optimal, achieving near-$\strongestVerifier{}$ accuracy at a fraction of the cost. 

We translate throughput to dollars using on-demand prices (January 22, 2026) and the hardware configurations in \autoref{sec:setup}. For CodeContests ($n{=}128$), costs per 100k programs form three tiers: majority voting baselines cost \oneTestMVCost{}--\tenTestMVCost{}, staged ORM verification costs \tenTestOrmCostMin{}--\tenTestOrmCostMax{}, and $\strongestVerifier{}$ costs \strongestVerifierCost{}. Staged verification is \stagedCostRatioMin{}--\stagedCostRatioMax{}$\times$ cheaper than $\strongestVerifier{}$ while recovering most of its accuracy. Moving from 1 to 10 tests adds only \oneTestTenTestCostDiff{} but gains \oneTestTenTestAccGain{} absolute accuracy.

{\setlength{\textfloatsep}{6pt}%
\begin{figure}[tb]
    \centering
    \includegraphics[width=1.0\columnwidth]{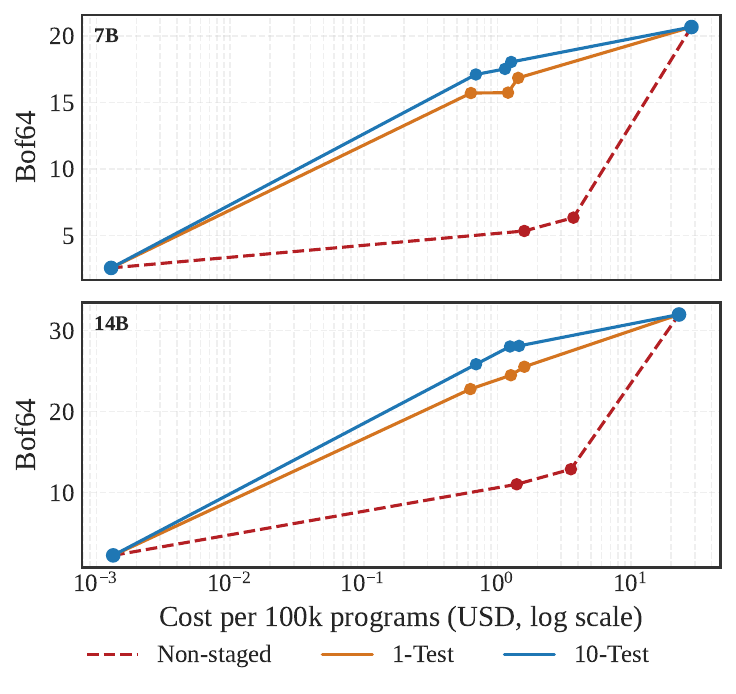}
    \caption{\textbf{Cost-accuracy Pareto frontier for Qwen 2.5 Coder Instruct models on CodeContests.}
  Top: 7B; bottom: 14B. Lines show Pareto curves for \textcolor[HTML]{b41f24}{non-staged},
  \textcolor[HTML]{d47420}{1-test}, and \textcolor[HTML]{1f77b4}{10-test}. Costs are per 100k programs (log scale).}
    \label{fig:cost-tradeoff-by-generator}
\end{figure}}

Costs are stable across generator sizes: verifier-only PPS ranges \oneTestPPSMin{}--\oneTestPPSMax{} for 1 test and \tenTestPPSMin{}--\tenTestPPSMax{} for 10 tests (\autoref{tab:codecontests-by-generator-results}). Cost estimates have tight 95\% CIs---the 10-test baseline costs \tenTestMVCost{} (CI: \tenTestMVCostCILow{}--\tenTestMVCostCIHigh{})---indicating throughput variance contributes negligibly to cost uncertainty compared to provider choice. \textbf{Staged verification is \stagedCostRatioMin{}--\stagedCostRatioMax{}$\times$ cheaper than full test-suite verification while recovering most of its accuracy, making it the Pareto-optimal choice for budget-constrained verification.}

%% file: sections/related_work.tex
\section{Related Work}
\textbf{Verification cost in code generation.} Execution-based evaluation is now standard in code benchmarks, from unit-test style tasks such as HumanEval and MBPP to larger suites such as APPS~\citep{chen_evaluating_2021,austin_program_2021,hendrycks_measuring_2021}. More recently, repository-scale settings such as SWE-Bench require environment setup and running a test suite per candidate, making verification a dominant cost in test-time scaling~\citep{jimenez_swe-bench_2024,Li2024DevEvalAM,Miserendino2025SWELancerCF,Orlanski_SlopCodeBench_Measuring_Code_2025}.

\textbf{Learned verifiers for selection.} Reward models, popularized by RLHF, are widely used for test-time selection (ORMs) and step-level evaluation (PRMs), and recent work also studies generative verification via LLM reasoning~\citep{stiennon_learning_2020,ouyang_training_2022,lightman_lets_2023,wang_math-shepherd_2024,zhang_generative_2024}. The closest work to ours is \citet{singhi_when_2025}, which studies when to sample more versus verify with an LLM; in contrast, we map the \emph{accuracy--throughput} frontier for code verification and study how execution filters and ORMs interact.

\textbf{Ranking programs.} A substantial line of work studies how to rank or rerank generated code candidates using execution signals, learned rankers, or auxiliary LLM-based reviewers~\citep{inala_fault-aware_2022,ni_lever_2023,sun_sifting_2024,zhang_coder_2022,zhong_debug_2024}. Compared to these works, we emphasize the compute dimension by reporting accuracy and throughput across verifiers, and analyze how lightweight execution filters complement neural rankers.

\textbf{Staged filtering and budgeted verification.} Staging verification---using cheap checks to reduce the number of candidates that reach more expensive validation---is common in software engineering workflows. Related ideas appear in automated program repair, where many candidate patches are generated and validated against test suites \citep{Jiang2023KNODDK}, and in work that reduces patch-validation cost or expensive test execution via prioritization and test selection~\citep{Benton2022Towards,Chen2020FastAP,Lou2021WhenAP,Yuan2022CIRCLE}. Similar cost--accuracy trade-offs also appear in multi-stage code generation systems that use cascades or budgeted selection~\citep{Chen2024ModelCF,ehrlich_codemonkeys_2025}. Our focus is on how combining lightweight execution filters with neural ranking shifts and optimizes the \emph{accuracy--throughput} Pareto frontier.

%% file: sections/conclusion.tex
\section{Conclusion}

Progress in verification for test-time scaling should be measured by Pareto improvement: moving the accuracy--throughput frontier outward. Our results show that the most consistent Pareto gains come from a simple staged verification pipeline of cheap filtering followed by ORM ranking---which achieves \tenTestPPSStrongImprove{} higher throughput with \tenTestAccStrongWorse{} lower accuracy than full test-suite verification. The need for the filter comes from a recurring ORM error mode: ORMs over-score near-miss programs whose tokens look right but fail on basic runtime semantics, such as division operator slips and mutation during iteration, placing filtered failures about \rankDifference{} ranks too high on average. Future work should target this failure mode directly and test whether the same frontier shifts appear when verification is dominated by much more substantial repository-scale overhead.

\input{sections/results_sections/limitations.tex}

%% file: sections/results_sections/limitations.tex
\paragraph{Limitations.}

While our motivation includes repository-scale verification such as SWE-Bench, our experiments focus on single-file Python benchmarks. We therefore do not measure costs that can dominate end-to-end verification at much larger scales, including container startup, dependency installation, and integration testing; we leave the evaluation of staged verification in such settings to future work.

Our generators and ORMs also share a common model family---Qwen---which may bias results. 
As additional powerful open weights models are released, the effects of varying the choice model family should be evaluated in future work. 
Additionally, results may differ under larger distribution shifts across languages, domains, or other aspects of the training pipeline. 
The dollar-cost analysis in \autoref{tab:cloud-costs} depends on specific hardware and cloud pricing assumptions; while these affect absolute costs, the empirical accuracy--throughput frontier is measured directly.

%% file: sections/appendix/train_data.tex
\section{Training Details Continued}\label{app:train-data}
\begin{table*}[h]
\centering
\caption{Details of our training datasets. ``\# Probs'' is the number of problems in the dataset. ``Prob. Tok'' is the average number of tokens in the problem description. ``Tokens'' is the average number of tokens in the problem and solutions when formatted with the prompt. ``Sols.'' is the average number of unique non-syntax error solutions per problem. ``\# Passed'' is the mean number of passing solutions per problem. ``\% Passed'' is the percentage of problems that had at least one passing solution. ``\# Pairs'' is the average number of possible pairs per problem, where a pair is 1 correct and 1 incorrect solution.}
\label{table:train-data}
\begin{tabular}{ll|rrrrrrr}
        Dataset & Split & \# Probs & Prob. Tok & Tokens & Sols. & \# Passed & \% Passed & \# Pairs \\
        \midrule
        \multirow[c]{2}{*}{CodeContests} & Train & 13,213 & 776.76 & 508.08 & 124.78 & 14.00 & 35.92 & 6.58 \\
         & Val & 117 & 921.66 & 625.66 & 126.31 & 4.36 & 23.08 & 3.61 \\
        \multirow[c]{2}{*}{GSM8K} & Train & 7,373 & 160.16 & 59.55 & 83.50 & 67.03 & 98.87 & 9.26 \\
         & Val & 100 & 177.42 & 66.63 & 105.93 & 78.66 & 100.00 & 15.16 \\
        \bottomrule
\end{tabular}
\end{table*}

\subsection{Hyperparameters}
We use The Transformers Library~\citep{wolf_huggingfaces_2020} with Hugging Face Datasets~\citep{lhoest_datasets_2021} for training our models. We use FlashAttention 2.0~\citep{dao_flashattention-2_2023} and BFloat16 precision. We use the Adam optimizer~\citep{kingma_adam_2017} with $\beta_1 = 0.9$, $\beta_2 = 0.999$, $\epsilon = 1e-8$. We use a cosine learning rate schedule that peaks at $5e-6$ and 10\% warmup steps. We use a batch size of 64 with a maximum sequence length of 2048. We employ gradient checkpointing and accumulation to be able to train our models on a single A6000 GPU. We train the ORM for 2 epochs on our dataset and use the last checkpoint for evaluation. We train each setup six times across six random seeds (1, 1999, 2000, 2021, 2024, 2042) and report the mean results. 
     
To create our training data we first apply Black formatting to the generated code as mentioned in \autoref{sec:setup}. We then filter out any programs with syntax errors and problems which do not have at least 1 correct and 1 incorrect solution. Then we randomly sample at most six examples per problem. For our pairwise objectives this means we sample a total of 12 programs per problem. We use the CodeContests validation set along with 100 randomly sampled problems from the GSM8K validation set to create our validation data.

%% file: sections/appendix/example_program.tex
\section{Example Problem}
\begin{lstlisting}[caption=Example Problem Description From CodeContests]
A string is called square if it is some string written twice in a row. For example, the strings "aa", "abcabc", "abab" and "baabaa" are square. But the strings "aaa", "abaaab" and "abcdabc" are not square.

For a given string s determine if it is square.

Input

The first line of input data contains an integer t (1 <= t <= 100) -the number of test cases.

This is followed by t lines, each containing a description of one test case. The given strings consist only of lowercase Latin letters and have lengths between 1 and 100 inclusive.

Output

For each test case, output on a separate line:

    * YES if the string in the corresponding test case is square, 
    * NO otherwise.

You can output YES and NO in any case (for example, strings yEs, yes, Yes and YES will be recognized as a positive response).

Example

Input

10
a
aa
aaa
aaaa
abab
abcabc
abacaba
xxyy
xyyx
xyxy

Output

NO
YES
NO
YES
YES
YES
NO
NO
NO
YES
\end{lstlisting}

\begin{lstlisting}[caption=Example Correct Solution,language=Python]
import sys

# Read input from stdin
input = sys.stdin.read
data = input().split()

# The first element is the number of test cases
t = int(data[0])

# Iterate over each test case
for i in range(1, t + 1):
    s = data[i]
    length = len(s)
    # Check if the length of the string is even
    if length % 2 == 0:
        half = length // 2
        # Check if the first half is equal to the second half
        if s[:half] == s[half:]:
            print("YES", end="\n")
        else:
            print("NO", end="\n")
    else:
        print("NO", end="\n")
\end{lstlisting}

\begin{lstlisting}[caption=Example Incorrect Solution,language=Python]
import sys

def is_square(s):
    n = len(s)
    for i in range(1, n // 2 + 1):
        if n % i == 0:
            prefix = s[:i]
            if prefix * (n // i) == s:
                return "YES"
    return "NO"

def main():
    input = sys.stdin.read
    data = input().split()
    t = int(data[0])
    results = []
    for i in range(1, t + 1):
        results.append(is_square(data[i]))
    for result in results:
        print(result)

if __name__ == "__main__":
    main()
\end{lstlisting}

%% file: sections/appendix/bug_examples.tex
\onecolumn
\section{Bug Pattern Examples}
\label{sec:bug-examples}

Examples of each bug pattern from \autoref{tab:bug-patterns-by-dataset}. All examples are from solutions ranked in the ORM's top-16 but filtered by $\weakVerifier{1}$.

\subsection{Float Division in Binary Search}
Using \texttt{/} instead of \texttt{//} produces floats that cause \texttt{TypeError} on indexing or silent truncation bugs. A single character difference.

\begin{lstlisting}[language=Python,caption={Wrong: float division (ORM Rank 2)}]
while right - left > 1e-9:
    mid = (left + right) / 2  # Bug: returns float 5.5, not int 5
    if check(k, mid, points):
        right = mid
    else:
        left = mid
\end{lstlisting}

\begin{lstlisting}[language=Python,caption={Correct: integer division}]
while left <= right:
    mid = (left + right) // 2  # Returns int 5
    if check(k, mid, points):
        right = mid - 1
    else:
        left = mid + 1
\end{lstlisting}

\subsection{Int Cast of Float Division}
Using \texttt{int(x/y)} instead of \texttt{x//y}. Appears correct but \texttt{int()} truncates toward zero while \texttt{//} floors, giving different results for negative numbers.

\begin{lstlisting}[language=Python,caption={Wrong: int of float division (ORM Rank 3)}]
prefix = [1]
for x, y in zip(a, b):
    if y == "*":
        prefix.append(prefix[-1] * x)
    else:
        prefix.append(int(prefix[-1] / x))  # Bug: int(-3/2)=-1, -3//2=-2
\end{lstlisting}

\begin{lstlisting}[language=Python,caption={Correct: floor division}]
        prefix.append(prefix[-1] // x)  # Correct floor division
\end{lstlisting}

\subsection{Large Integer Arithmetic}
Operations on large integers that lose precision or overflow intermediate values. Common in competitive programming where results require modular arithmetic.

\begin{lstlisting}[language=Python,caption={Wrong: intermediate overflow (ORM Rank 3)}]
def count_paths(n, edges):
    MOD = 10**9 + 7
    # Bug: a * b overflows before modulo on some paths
    result = a * b * c % MOD  # Intermediate a*b*c exceeds int precision
    return result
\end{lstlisting}

\begin{lstlisting}[language=Python,caption={Correct: modulo at each step}]
def count_paths(n, edges):
    MOD = 10**9 + 7
    result = a * b % MOD
    result = result * c % MOD  # Keep intermediate values bounded
    return result
\end{lstlisting}

\subsection{Remove in Loop}
Calling \texttt{.remove()} inside a loop modifies the collection while iterating, causing skipped elements or \texttt{ValueError}.

\begin{lstlisting}[language=Python,caption={Wrong: modifying set while iterating (ORM Rank 4)}]
for i in range(1, n):
    if (u, i) in E and (v, i) not in E:
        T = min(T, f(u, i, t + 1))
        E.remove((u, i))  # Bug: modifies E during iteration
        E.remove((v, i))
        E.add((u, v))
\end{lstlisting}

\begin{lstlisting}[language=Python,caption={Correct: collect then remove}]
to_remove = []
for i in range(1, n):
    if (u, i) in E and (v, i) not in E:
        T = min(T, f(u, i, t + 1))
        to_remove.append(((u, i), (v, i)))
for pair in to_remove:
    E.discard(pair[0])
    E.discard(pair[1])
\end{lstlisting}

\subsection{While Len Condition}
Using \texttt{while len(x) > 0} instead of \texttt{while x}. Verbose and often indicates confused reasoning about the loop invariant.

\begin{lstlisting}[language=Python,caption={Wrong: verbose length check (ORM Rank 5)}]
while len(habitats) > 0:  # Unidiomatic
    if is_inside((center, radius), habitats[-1]):
        habitats.pop()
    else:
        radius += dist((center, radius), habitats[-1])
        # Bug: infinite loop if habitat can't be covered
\end{lstlisting}

\begin{lstlisting}[language=Python,caption={Correct: idiomatic check with termination}]
while habitats:
    habitat = habitats[-1]
    if is_inside((center, radius), habitat):
        habitats.pop()
    elif radius >= max_radius:
        break  # Explicit termination condition
    else:
        radius = compute_new_radius(center, habitat)
\end{lstlisting}

%% file: sections/appendix/prompts.tex
\section{Prompts Used}\label{app:prompts}
\begin{lstlisting}[caption=Prompt used for training and inference]
# Question
Tom and Tim both brought 4, six-sided dice to school. 
How many total sides are there?

# Proposed Solution
```python
def solution():
    dice_per_person = 4
    sides_per_die = 6
    total_dice = dice_per_person * 2
    total_sides = total_dice * sides_per_die
    result = total_sides
    return result
    
```
<|end_of_text|>
\end{lstlisting}

%% file: sections/appendix/execution_details.tex
\section{Execution Details Continued}\label{app:execution-details}

HumanEval tests function completion where the model must write the correct function given a docstring and signature. MBPP, on the other hand, tasks the model with synthesizing a function given a single sentence specification and a single test case. Then we use the execution framework provided by EvalPlus that executes the program from inside of a Python process. In both of these cases, the evaluation relies on assertions that call a specific entrypoint function. Thus, we use EvalPlus {\tt sanitize} method to extract all relevant code and comments using treesitter\footnote{\url{https://github.com/tree-sitter/tree-sitter}} to parse the AST. This does come with the caveat that it heavily perturbs the whitespace of the original code in unnatural ways. Thus we use the Black formatter\footnote{\url{https://github.com/psf/black}} to format both our evaluation and training data. We run all of our trials on a cluster designed for high throughput computing~\citep{chtc}.

%% file: sections/appendix/data.tex
\section{Data Generation Details}\label{app:data-generation}
To generate our training data for our ORMs, we use the training splits of CodeContests-Python~\citep{li_competition-level_2022} and  GSM8K~\citep{cobbe_training_2021} in Program-aided Language Models (PAL) format~\citep{gao_pal_2022}. The CodeContests dataset is a collection of competitive programming problems from platforms such as CodeForces.\footnote{\url{https://codeforces.com/}} Each program requires taking in STDIN inputs and printing the results to STDOUT. Most, but not all, problems have a few private test cases. We follow the LiveCodeBench~\citep{jain_livecodebench_2024} prompts and formatting for generation. 
For GSM8K, we opt for the PAL format as it allows us to have more programming training data from a different domain. We additionally use the BigCodeEval~\citep{allal_bigcode-projectbigcode-evaluation-harness_2025} in context examples for GSM8K. We use Qwen 2.5 Coder Instruct 7B~\citep{hui_qwen25-coder_2024} with Int8 GPTQ quantization~\citep{frantar_gptq_2023} to generate our training data. We sample a maximum of 1024 tokens $\trainNumSampled$ times per problem with a temperature of $T=1.0$ and top-p of $0.95$. We use the VLLM~\citep{kwon_efficient_2023} library to generate our data. Full details of our training data are found in \autoref{app:train-data}.

%% file: sections/appendix/eval_data.tex
\begin{table*}
    \centering
    \caption{Breakdown of evaluation data. $\generator$ is the generator size, $T$ is the temperature, and DS is the dataset. Prob Pass is the percentage of problems where at least one solution passes at least one test. \% Pass is the mean fraction of tests passed per solution. Net Sols is the total number of solutions. Prob. Tok and Tok. are the average token counts for problem statements and solutions. \# Pass is the average number of fully correct solutions (passing all tests) per problem. Avg. Sols is the average number of solutions per problem.}\label{tab:eval-data}
    \resizebox{\textwidth}{!}{%
    \begin{tabular}{lll|ccccccc}
        \toprule
        $\generator$ & $T$ & DS  & Prob Pass & \% Pass & Net Sols & Prob. Tok & Tok. & \# Pass & Avg. Sols \\
        \midrule
        \multirow[c]{4}{*}{500M} & \multirow[c]{4}{*}{1.0} & CC & 68.48 & 2.75 & 21,014 & 620.74 & 878.31 & 0.04 & 127.36 \\
         &  & GSM8K & 80.36 & 18.75 & 151,608 & 71.31 & 155.68 & 18.31 & 114.94 \\
         &  & HE & 100.00 & 41.99 & 16,444 & 144.22 & 359.51 & 33.98 & 100.27 \\
         &  & MBPP & 95.77 & 40.42 & 40,187 & 60.51 & 156.05 & 39.50 & 106.31 \\
        \midrule
        \multirow[c]{4}{*}{1.5B} & \multirow[c]{4}{*}{1.0} & CC & 78.18 & 6.03 & 20,932 & 620.74 & 868.14 & 0.13 & 126.86 \\
         &  & GSM8K & 95.00 & 50.33 & 135,144 & 71.31 & 160.51 & 45.92 & 102.46 \\
         &  & HE & 100.00 & 57.90 & 16,643 & 144.22 & 320.82 & 51.06 & 101.48 \\
         &  & MBPP & 98.41 & 54.72 & 41,018 & 60.51 & 132.90 & 53.15 & 108.51 \\
        \midrule
        \multirow[c]{4}{*}{3B} & \multirow[c]{4}{*}{1.0} & CC & 88.48 & 9.83 & 21,108 & 620.74 & 921.48 & 0.47 & 127.93 \\
         &  & GSM8K & 97.80 & 66.16 & 129,375 & 71.31 & 163.22 & 59.09 & 98.09 \\
         &  & HE & 100.00 & 75.27 & 14,935 & 144.22 & 336.42 & 63.43 & 91.07 \\
         &  & MBPP & 99.47 & 58.99 & 44,896 & 60.51 & 138.26 & 65.64 & 118.77 \\
        \midrule
        \multirow[c]{20}{*}{7B} & \multirow[c]{4}{*}{0.2} & CC & 84.24 & 14.54 & 11,905 & 620.74 & 902.13 & 0.21 & 72.15 \\
         &  & GSM8K & 91.66 & 82.20 & 14,135 & 71.31 & 162.98 & 7.10 & 10.72 \\
         &  & HE & 99.39 & 86.09 & 2,643 & 144.22 & 323.89 & 12.14 & 16.12 \\
         &  & MBPP & 96.83 & 72.24 & 10,281 & 60.51 & 144.11 & 14.60 & 27.20 \\
        \cline{2-10}
         & \multirow[c]{4}{*}{0.4} & CC & 88.48 & 14.92 & 18,331 & 620.74 & 895.37 & 0.32 & 111.10 \\
         &  & GSM8K & 95.68 & 82.11 & 40,932 & 71.31 & 163.93 & 21.26 & 31.03 \\
         &  & HE & 99.39 & 83.88 & 7,071 & 144.22 & 324.28 & 33.40 & 43.12 \\
         &  & MBPP & 97.88 & 71.08 & 24,167 & 60.51 & 144.51 & 37.24 & 63.93 \\
        \cline{2-10}
         & \multirow[c]{4}{*}{0.6} & CC & 90.30 & 14.80 & 20,300 & 620.74 & 892.90 & 0.42 & 123.03 \\
         &  & GSM8K & 96.89 & 81.23 & 69,711 & 71.31 & 164.16 & 37.52 & 52.85 \\
         &  & HE & 99.39 & 83.23 & 11,127 & 144.22 & 319.59 & 52.44 & 67.85 \\
         &  & MBPP & 98.94 & 69.65 & 34,328 & 60.51 & 144.02 & 55.72 & 90.81 \\
        \cline{2-10}
         & \multirow[c]{4}{*}{0.8} & CC & 93.33 & 14.47 & 20,834 & 620.74 & 893.17 & 0.38 & 126.27 \\
         &  & GSM8K & 97.88 & 80.62 & 94,225 & 71.31 & 164.30 & 52.23 & 71.44 \\
         &  & HE & 99.39 & 81.76 & 14,241 & 144.22 & 315.55 & 67.17 & 86.84 \\
         &  & MBPP & 99.21 & 68.69 & 40,961 & 60.51 & 143.52 & 68.29 & 108.36 \\
        \cline{2-10}
         & \multirow[c]{4}{*}{1.0} & CC & 92.12 & 14.11 & 21,022 & 620.74 & 897.16 & 0.32 & 127.41 \\
         &  & GSM8K & 98.56 & 79.44 & 114,882 & 71.31 & 165.06 & 64.48 & 87.10 \\
         &  & HE & 99.39 & 80.98 & 16,625 & 144.22 & 310.11 & 78.46 & 101.37 \\
         &  & MBPP & 99.74 & 66.64 & 44,915 & 60.51 & 144.73 & 74.42 & 118.82 \\
         \midrule
        \multirow[c]{4}{*}{14B} & \multirow[c]{4}{*}{1.0} & CC & 92.73 & 18.76 & 19,852 & 620.74 & 988.10 & 0.10 & 120.32 \\
         &  & GSM8K & 98.94 & 87.69 & 129,548 & 71.31 & 170.84 & 83.86 & 98.22 \\
         &  & HE & 100.00 & 83.97 & 15,410 & 144.22 & 314.70 & 75.10 & 93.96 \\
         &  & MBPP & 98.94 & 70.23 & 24,411 & 60.51 & 142.26 & 38.29 & 64.58 \\
        \bottomrule
    \end{tabular}
    }
\end{table*}

%% file: sections/appendix/scale_graphs.tex
\begin{figure}[h]
    \centering
    \includegraphics[width=1\textwidth]{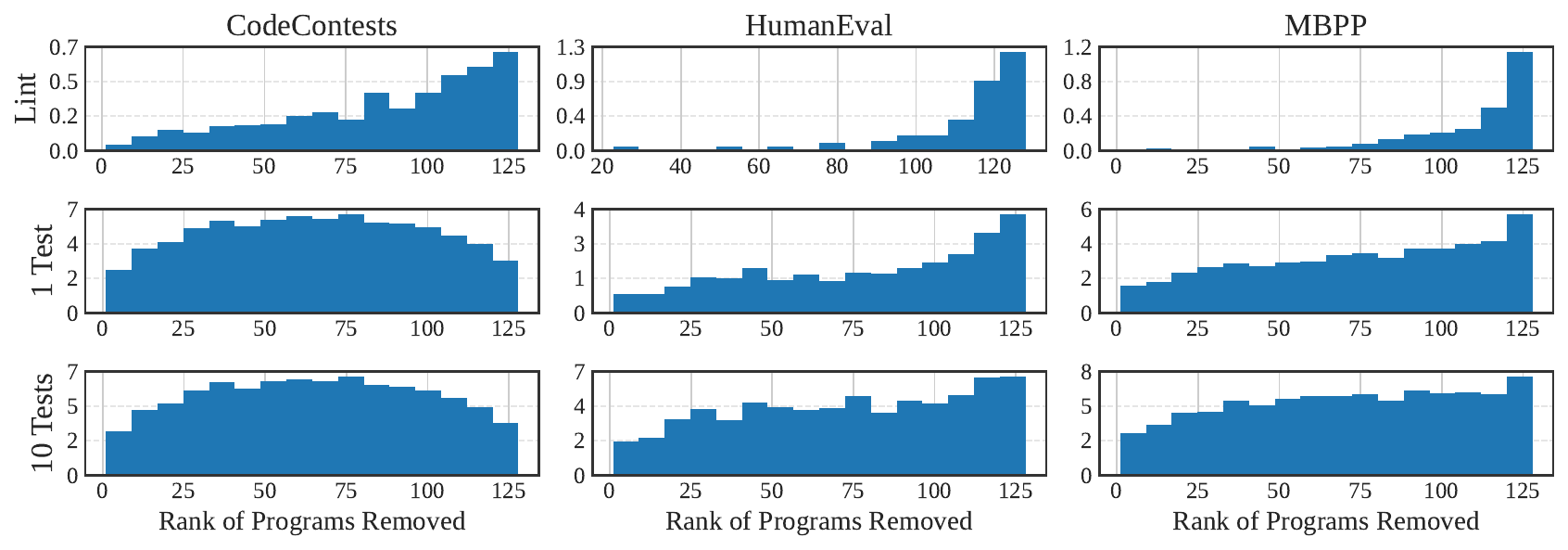}
    \caption{Distribution of the failed candidates based on where $\ormVerifier{}$ had ranked them. A rank of 1 means the candidate was the top-ranked, while 128 means it was the lowest-ranked. The rows are the individual verifiers while the columns are the datasets. This is for the 500M $\ormVerifier$.}
    \label{fig:true-ranking-qc-inst-500m-t10-n128}
\end{figure}

%% file: tables/results_std_table.tex
\begin{table*}
    \centering

    \caption{Standard deviation of the results for the different pruning with a weak verifier then ranking methods.  {\colorbox{textgreen}{Green backgrounds}} is higher performance while {\colorbox{textred}{Red backgrounds}} is lower performance with respect to the entire column. If $\verifier{}$ is ``---'' that means no pruning is done. $\strongestVerifier{}$ is the case where \emph{all} test cases are run. ``Syntax'' and ``Lint'' remove any programs with the respective errors. ``N Test'' prunes out any programs that do not pass the first $N$ test cases. The evaluation dataset is generated with Qwen 2.5 Coder 7B Instruct using $T=1.0$, $n=128$, $top_p=0.95$, and 1024 tokens. For the Non-ORM methods, there is no standard deviation for best-of-64 as the ordering is the same} \label{tab:overall-std-results}
    \resizebox{0.9\textwidth}{!}{%
    \begin{tabular}{ll|cc|cc|cc|cc}
\toprule
 &  & \multicolumn{2}{c|}{CodeContests} & \multicolumn{2}{c|}{GSM8K} & \multicolumn{2}{c|}{HumanEval} & \multicolumn{2}{c}{MBPP} \\ Filter & Ranker & Bof64 & PPS & Bof64 & PPS & Bof64 & PPS & Bof64 & PPS \\
\midrule
\multirow[c]{3}{*}{---} & MV & --- & --- & --- & --- & --- & --- & --- & --- \\
 & 500M & 0.74 & 14.45 & 0.99 & 71.70 & 1.05 & 37.57 & 1.06 & 76.28 \\
 & 1.5B & 0.69 & 4.00 & 1.28 & 20.18 & 1.54 & 10.92 & 1.20 & 23.13 \\
\midrule
\multirow[c]{3}{*}{Syntax} & MV & --- & 399.40 & --- & 445.70 & --- & 517.91 & --- & 742.06 \\
 & 500M & 0.70 & 2.73 & 0.30 & 13.23 & 0.93 & 5.26 & 0.81 & 15.19 \\
 & 1.5B & 0.76 & 1.76 & 0.43 & 8.23 & 1.80 & 4.19 & 0.83 & 9.72 \\
\midrule
\multirow[c]{3}{*}{Lint} & MV & --- & 92.07 & --- & 9.88 & --- & 4.69 & --- & 9.62 \\
 & 500M & 0.87 & 9.60 & 1.26 & 18.62 & 0.89 & 12.14 & 2.26 & 11.73 \\
 & 1.5B & 0.82 & 2.75 & 1.98 & 7.56 & 1.41 & 5.02 & 1.74 & 7.12 \\
\midrule
\multirow[c]{3}{*}{1 Test} & MV & --- & 1.48 & --- & --- & --- & 101.45 & --- & 150.46 \\
 & 500M & 0.34 & 0.28 & --- & --- & 1.03 & 27.92 & 1.42 & 48.73 \\
 & 1.5B & 0.49 & 0.61 & --- & --- & 1.54 & 9.17 & 1.00 & 20.22 \\
\midrule
\multirow[c]{3}{*}{10 Tests} & MV & --- & 2.27 & --- & --- & --- & 54.43 & --- & 44.48 \\
 & 500M & 0.15 & 0.01 & --- & --- & 0.76 & 0.27 & 0.39 & 0.03 \\
 & 1.5B & 0.49 & 0.04 & --- & --- & 0.82 & 0.06 & 0.18 & 0.08 \\
\midrule
\multicolumn{2}{c|}{$\strongestVerifier{}$} & --- & 0.09 & --- & 37.90 & --- & 0.92 & --- & 0.43 \\
\bottomrule
\end{tabular}

    }
\end{table*}

%% file: sections/appendix/tables.tex
\section{All result tables}\label{app:all-tables-that-are-tabular}
\input{tables/cloud_costs.tex}
\input{tables/pruning_tokens.tex}
\input{tables/filtered_sols.tex}
\input{tables/scale/results_table_500M_1_0.tex}
\input{tables/scale/results_table_1.5B_1_0.tex}
\input{tables/scale/results_table_3B_1_0.tex}
\input{tables/scale/results_table_14B_1_0.tex}
\input{tables/scale/results_table_7B_0_2.tex}
\input{tables/scale/results_table_7B_0_4.tex}
\input{tables/scale/results_table_7B_0_6.tex}
\input{tables/scale/results_table_7B_0_8.tex}

%% file: tables/cloud_costs.tex
\begin{table}[t]
    \centering
    \caption{On-demand hourly prices for hardware-matched instances. Prices are Linux on-demand in AWS us-east-1, Azure eastus, and GCP us-central1 (Jan 22, 2026).} \label{tab:cloud-costs}
    \small
    \begin{tabular}{lcc}
\toprule
Provider& GPU (single $\sim$48GB) & CPU 64c/256GB \\
\midrule
AWS  & \$3.004/hr & \$3.072/hr \\
Azure  & \$3.673/hr & \$3.072/hr \\
GCP  & \$5.782/hr & \$3.1078/hr \\
\bottomrule
    \end{tabular}
\end{table}

%% file: tables/pruning_tokens.tex
\begin{table*}
    \centering
    \caption{Information about the tokens filtered from the datasets by the 5 verifiers we look at on the 3B, 7B, and 14B models. The percentage filtered is the percentage of tokens filtered from the dataset, the average sequence length is the average sequence length of the tokens that were filtered, and the total tokens filtered is the total number of tokens filtered from the dataset. }\label{tab:pruning_tokens}
    \resizebox{0.8\textwidth}{!}{%
    \begin{tabular}{lll|rrr}
        \toprule
        $\generator$ & $V$ & DS & \% Filtered & Avg. Seq Length & Total Tokens Filtered \\
       \midrule
       \multirow[c]{17}{*}{7B} & \multirow[c]{4}{*}{Syntax} & CC & 0.76 & 1,222.81 & 1.89e+07 \\
        &  & GSM8K & 0.45 & 201.33 & 2.04e+07 \\
        &  & HE & 0.00 & 135.00 & 5.47e+06 \\
        &  & MBPP & 0.00 & 65.00 & 6.63e+06 \\
        \cline{2-6}
        & \multirow[c]{4}{*}{Lint} & CC & 5.36 & 1,131.51 & 1.89e+07 \\
        &  & GSM8K & 1.33 & 203.86 & 2.04e+07 \\
        &  & HE & 1.00 & 408.92 & 5.47e+06 \\
        &  & MBPP & 1.05 & 159.68 & 6.63e+06 \\
        \cline{2-6}
        & \multirow[c]{3}{*}{1 Test} & CC & 88.90 & 919.54 & 1.89e+07 \\
        &  & HE & 8.44 & 378.60 & 5.47e+06 \\
        &  & MBPP & 22.88 & 173.66 & 6.63e+06 \\
        \cline{2-6}
        & \multirow[c]{3}{*}{3 Tests} & CC & 92.85 & 908.04 & 1.89e+07 \\
        &  & HE & 13.97 & 364.37 & 5.47e+06 \\
        &  & MBPP & 27.58 & 170.55 & 6.63e+06 \\
        \cline{2-6}
        & \multirow[c]{3}{*}{10 Tests} & CC & 93.39 & 906.89 & 1.89e+07 \\
        &  & HE & 18.84 & 364.83 & 5.47e+06 \\
        &  & MBPP & 37.92 & 165.21 & 6.63e+06 \\
        \midrule
       \multirow[c]{17}{*}{14B} & \multirow[c]{4}{*}{Syntax} & CC & 4.87 & 1,610.94 & 1.96e+07 \\
        &  & GSM8K & 0.16 & 213.89 & 2.35e+07 \\
        &  & HE & 0.04 & 130.29 & 5.22e+06 \\
        &  & MBPP & 0.13 & 61.49 & 3.93e+06 \\
        \cline{2-6}
        & \multirow[c]{4}{*}{Lint} & CC & 7.76 & 1,411.36 & 1.96e+07 \\
        &  & GSM8K & 0.81 & 216.55 & 2.35e+07 \\
        &  & HE & 0.95 & 315.97 & 5.22e+06 \\
        &  & MBPP & 0.94 & 135.12 & 3.93e+06 \\
        \cline{2-6}
        & \multirow[c]{3}{*}{1 Test} & CC & 85.08 & 1,019.14 & 1.96e+07 \\
        &  & HE & 5.99 & 375.57 & 5.22e+06 \\
        &  & MBPP & 21.93 & 182.11 & 3.93e+06 \\
        \cline{2-6}
        & \multirow[c]{3}{*}{3 Tests} & CC & 89.01 & 1,005.72 & 1.96e+07 \\
        &  & HE & 11.18 & 369.63 & 5.22e+06 \\
        &  & MBPP & 26.60 & 178.78 & 3.93e+06 \\
        \cline{2-6}
        & \multirow[c]{3}{*}{10 Tests} & CC & 89.95 & 1,003.10 & 1.96e+07 \\
        &  & HE & 15.53 & 373.36 & 5.22e+06 \\
        &  & MBPP & 38.54 & 173.15 & 3.93e+06 \\
        \bottomrule
       \end{tabular}
    }
\end{table*}

%% file: tables/filtered_sols.tex
\begin{table*}
    \centering
    \caption{Information on filtered solutions for the 500M and 1.5B $\ormVerifier$ across the datasets when using the 7B Qwen 2.5 Coder Instruct $\generator$ and a temperature of 1.0. ``\# Rem'' is the average number of solutions removed per question. ``Avg. Rank'' is the average rank of the solutions that were removed. ``M1 Rank'' is the average highest rank of the solutions that were removed. ``M5 Rank'' is the average rank of the 5 top ranks of the solutions that were removed. ``\% Removed'' is the total percentage of solutions removed across all questions. ``\% Prob'' is the percentage of problems where all solutions were removed. ``Spread'' is the average spread of the solutions that were removed across the six seeds.}
    \label{tab:filter-res}
    \resizebox{\textwidth}{!}{%
    \begin{tabular}{ll|rrrrrrr}
        $\ormVerifier$ & $V$ & \# Rem & Avg. Rank & M1 Rank & M5 Rank & \% Removed & \% Prob & Spread \\
        \midrule
       \multirow[c]{4}{*}{500M} & 1 Test & 51.38 & 82.90 & 38.87 & 51.04 & 40.14 & 6.86 & 20.00 \\
        & 10 Tests & 63.53 & 77.81 & 27.63 & 39.36 & 49.63 & 12.57 & 17.90 \\
        & Lint & 4.76 & 102.31 & 86.49 & 97.84 & 3.72 & 0.00 & 31.91 \\
        & Syntax & 5.13 & 106.50 & 98.53 & 104.93 & 4.01 & 0.00 & 32.08 \\
        \midrule
       \multirow[c]{4}{*}{1.5B} & 1 Test & 51.38 & 89.27 & 47.87 & 58.42 & 40.14 & 6.86 & 19.14 \\
       & 10 Tests & 63.53 & 83.47 & 35.41 & 45.89 & 49.63 & 12.57 & 17.14 \\
       & Lint & 4.76 & 108.70 & 96.02 & 104.91 & 3.72 & 0.00 & 30.16 \\
       & Syntax & 5.13 & 112.99 & 107.00 & 111.73 & 4.01 & 0.00 & 30.28 \\
        \bottomrule
       \end{tabular}
    }
\end{table*}

%% file: tables/scale/results_table_500M_1_0.tex
\begin{table*}
    \centering
    
    \caption{Results for the generation with Qwen 2.5 Coder 500M instruct with $T=1.0$, $n=128$, $top_p=0.95$, and 1024 tokens.  {\colorbox{textgreen}{Green backgrounds}} is higher performance while {\colorbox{textred}{Red backgrounds}} is lower performance with respect to the entire column. If $\verifier{}$ is ``---'' that means no pruning is done. $\strongestVerifier{}$ is the case where \emph{all} test cases are run.}\label{tab:500M-1.0-overall-results}
    \resizebox{0.9\textwidth}{!}{%
    \begin{tabular}{ll|cc|cc|cc|cc}
\toprule
 &  & \multicolumn{2}{c|}{CodeContests} & \multicolumn{2}{c|}{GSM8K} & \multicolumn{2}{c|}{HumanEval} & \multicolumn{2}{c}{MBPP} \\ Filter & Ranker & Bof64 & PPS & Bof64 & PPS & Bof64 & PPS & Bof64 & PPS \\
\midrule
\multirow[c]{2}{*}{---} & 500M & \cellcolor[rgb]{0.722,0.510,0.510} 0.47 & \cellcolor[rgb]{0.853,0.741,0.741} 68.39 & \cellcolor[rgb]{0.722,0.510,0.510} 38.08 & \cellcolor[rgb]{0.510,0.722,0.569} 374.52 & \cellcolor[rgb]{0.722,0.510,0.510} 51.96 & \cellcolor[rgb]{0.950,0.913,0.913} 157.37 & \cellcolor[rgb]{0.722,0.510,0.510} 47.70 & \cellcolor[rgb]{0.831,0.904,0.852} 353.18 \\
 & 1.5B & \cellcolor[rgb]{0.743,0.547,0.547} 0.61 & \cellcolor[rgb]{0.722,0.510,0.510} 26.42 & \cellcolor[rgb]{0.510,0.722,0.569} 53.91 & \cellcolor[rgb]{0.722,0.510,0.510} 147.74 & \cellcolor[rgb]{0.879,0.787,0.787} 67.50 & \cellcolor[rgb]{0.722,0.510,0.510} 64.10 & \cellcolor[rgb]{0.820,0.682,0.682} 57.70 & \cellcolor[rgb]{0.722,0.510,0.510} 145.72 \\
\midrule
\multirow[c]{3}{*}{1 Test} & MV & \cellcolor[rgb]{0.861,0.755,0.755} 1.39 & \cellcolor[rgb]{0.510,0.722,0.569} 126.19 & --- & --- & \cellcolor[rgb]{0.837,0.714,0.714} 64.05 & \cellcolor[rgb]{0.510,0.722,0.569} 851.41 & \cellcolor[rgb]{0.856,0.746,0.746} 61.41 & \cellcolor[rgb]{0.510,0.722,0.569} 870.83 \\
 & 500M & \cellcolor[rgb]{1.000,1.000,1.000} 1.71 & \cellcolor[rgb]{1.000,1.000,1.000} 79.64 & --- & --- & \cellcolor[rgb]{0.851,0.738,0.738} 65.51 & \cellcolor[rgb]{1.000,1.000,1.000} 180.55 & \cellcolor[rgb]{0.882,0.792,0.792} 62.42 & \cellcolor[rgb]{1.000,1.000,1.000} 325.07 \\
 & 1.5B & \cellcolor[rgb]{0.632,0.791,0.676} 2.02 & \cellcolor[rgb]{0.914,0.848,0.848} 74.27 & --- & --- & \cellcolor[rgb]{1.000,1.000,1.000} 74.18 & \cellcolor[rgb]{0.796,0.641,0.641} 91.64 & \cellcolor[rgb]{0.919,0.954,0.928} 66.59 & \cellcolor[rgb]{0.861,0.756,0.756} 200.19 \\
\midrule
\multirow[c]{3}{*}{3 Tests} & MV & \cellcolor[rgb]{0.861,0.755,0.755} 1.39 & \cellcolor[rgb]{0.516,0.725,0.574} 125.72 & --- & --- & \cellcolor[rgb]{0.963,0.935,0.935} 72.13 & \cellcolor[rgb]{0.531,0.733,0.587} 814.36 & \cellcolor[rgb]{0.966,0.940,0.940} 64.44 & \cellcolor[rgb]{0.698,0.828,0.734} 483.11 \\
 & 500M & \cellcolor[rgb]{0.986,0.992,0.988} 1.72 & \cellcolor[rgb]{0.891,0.808,0.808} 72.82 & --- & --- & \cellcolor[rgb]{0.990,0.994,0.991} 74.46 & \cellcolor[rgb]{0.964,0.980,0.968} 215.28 & \cellcolor[rgb]{1.000,1.000,1.000} 65.26 & \cellcolor[rgb]{0.937,0.964,0.944} 335.60 \\
 & 1.5B & \cellcolor[rgb]{0.604,0.775,0.652} 2.05 & \cellcolor[rgb]{0.855,0.745,0.745} 69.09 & --- & --- & \cellcolor[rgb]{0.832,0.904,0.852} 78.73 & \cellcolor[rgb]{0.862,0.757,0.757} 116.20 & \cellcolor[rgb]{0.839,0.908,0.858} 67.90 & \cellcolor[rgb]{0.876,0.781,0.781} 213.38 \\
\midrule
\multirow[c]{3}{*}{10 Tests} & MV & \cellcolor[rgb]{0.974,0.954,0.954} 1.65 & \cellcolor[rgb]{0.517,0.725,0.575} 125.65 & --- & --- & \cellcolor[rgb]{0.577,0.760,0.628} 82.88 & \cellcolor[rgb]{0.640,0.796,0.684} 620.28 & \cellcolor[rgb]{0.646,0.799,0.689} 70.64 & \cellcolor[rgb]{0.749,0.857,0.779} 378.63 \\
 & 500M & \cellcolor[rgb]{0.932,0.961,0.940} 1.76 & \cellcolor[rgb]{0.929,0.960,0.937} 87.47 & --- & --- & \cellcolor[rgb]{0.549,0.744,0.603} 83.21 & \cellcolor[rgb]{0.998,0.999,0.998} 182.60 & \cellcolor[rgb]{0.567,0.754,0.619} 71.64 & \cellcolor[rgb]{0.860,0.753,0.753} 199.45 \\
 & 1.5B & \cellcolor[rgb]{0.510,0.722,0.569} 2.15 & \cellcolor[rgb]{0.976,0.986,0.978} 82.33 & --- & --- & \cellcolor[rgb]{0.510,0.722,0.569} 83.67 & \cellcolor[rgb]{0.859,0.752,0.752} 114.91 & \cellcolor[rgb]{0.510,0.722,0.569} 72.36 & \cellcolor[rgb]{0.744,0.550,0.550} 154.63 \\
\midrule
\multicolumn{2}{c|}{$\strongestVerifier{}$} & \cellcolor[rgb]{0.510,0.722,0.569} 2.16 & \cellcolor[rgb]{0.722,0.510,0.510} 3.32 & \cellcolor[rgb]{0.510,0.722,0.569} 73.63 & \cellcolor[rgb]{0.722,0.510,0.510} 391.53 & \cellcolor[rgb]{0.510,0.722,0.569} 86.79 & \cellcolor[rgb]{0.722,0.510,0.510} 19.09 & \cellcolor[rgb]{0.510,0.722,0.569} 73.73 & \cellcolor[rgb]{0.722,0.510,0.510} 13.99 \\
\bottomrule
\end{tabular}

    }
\end{table*}

%% file: tables/scale/results_table_1.5B_1_0.tex
\begin{table*}
    \centering
    
    \caption{Results for the generation with Qwen 2.5 Coder 1.5B instruct with $T=1.0$, $n=128$, $top_p=0.95$, and 1024 tokens.  {\colorbox{textgreen}{Green backgrounds}} is higher performance while {\colorbox{textred}{Red backgrounds}} is lower performance with respect to the entire column. If $\verifier{}$ is ``---'' that means no pruning is done. $\strongestVerifier{}$ is the case where \emph{all} test cases are run.}\label{tab:1.5B-1.0-overall-results}
    \resizebox{0.9\textwidth}{!}{%
    \begin{tabular}{ll|cc|cc|cc|cc}
\toprule
 &  & \multicolumn{2}{c|}{CodeContests} & \multicolumn{2}{c|}{GSM8K} & \multicolumn{2}{c|}{HumanEval} & \multicolumn{2}{c}{MBPP} \\ Filter & Ranker & Bof64 & PPS & Bof64 & PPS & Bof64 & PPS & Bof64 & PPS \\
\midrule
\multirow[c]{2}{*}{---} & 500M & \cellcolor[rgb]{0.722,0.510,0.510} 1.64 & \cellcolor[rgb]{0.852,0.739,0.739} 67.79 & \cellcolor[rgb]{0.722,0.510,0.510} 63.63 & \cellcolor[rgb]{0.510,0.722,0.569} 348.16 & \cellcolor[rgb]{0.722,0.510,0.510} 64.78 & \cellcolor[rgb]{0.992,0.995,0.993} 169.35 & \cellcolor[rgb]{0.722,0.510,0.510} 58.72 & \cellcolor[rgb]{0.857,0.919,0.875} 394.16 \\
 & 1.5B & \cellcolor[rgb]{0.760,0.578,0.578} 2.62 & \cellcolor[rgb]{0.722,0.510,0.510} 26.63 & \cellcolor[rgb]{0.510,0.722,0.569} 75.01 & \cellcolor[rgb]{0.722,0.510,0.510} 137.80 & \cellcolor[rgb]{0.847,0.731,0.731} 71.85 & \cellcolor[rgb]{0.722,0.510,0.510} 69.56 & \cellcolor[rgb]{0.792,0.634,0.634} 65.72 & \cellcolor[rgb]{0.784,0.620,0.620} 162.73 \\
\midrule
\multirow[c]{3}{*}{1 Test} & MV & \cellcolor[rgb]{0.851,0.739,0.739} 4.92 & \cellcolor[rgb]{0.510,0.722,0.569} 124.94 & --- & --- & \cellcolor[rgb]{0.853,0.741,0.741} 72.17 & \cellcolor[rgb]{0.555,0.747,0.609} 905.65 & \cellcolor[rgb]{0.874,0.777,0.777} 72.81 & \cellcolor[rgb]{0.510,0.722,0.569} 1195.51 \\
 & 500M & \cellcolor[rgb]{0.943,0.968,0.950} 5.64 & \cellcolor[rgb]{1.000,1.000,1.000} 74.06 & --- & --- & \cellcolor[rgb]{0.869,0.770,0.770} 73.05 & \cellcolor[rgb]{1.000,1.000,1.000} 162.21 & \cellcolor[rgb]{0.858,0.751,0.751} 72.34 & \cellcolor[rgb]{1.000,1.000,1.000} 312.70 \\
 & 1.5B & \cellcolor[rgb]{0.997,0.995,0.995} 5.60 & \cellcolor[rgb]{0.850,0.737,0.737} 67.36 & --- & --- & \cellcolor[rgb]{0.961,0.931,0.931} 77.90 & \cellcolor[rgb]{0.790,0.630,0.630} 80.87 & \cellcolor[rgb]{0.952,0.916,0.916} 74.26 & \cellcolor[rgb]{0.855,0.746,0.746} 186.30 \\
\midrule
\multirow[c]{3}{*}{3 Tests} & MV & \cellcolor[rgb]{0.933,0.881,0.881} 5.39 & \cellcolor[rgb]{0.530,0.733,0.586} 122.97 & --- & --- & \cellcolor[rgb]{0.994,0.997,0.995} 80.10 & \cellcolor[rgb]{0.510,0.722,0.569} 1026.14 & \cellcolor[rgb]{0.930,0.960,0.939} 76.22 & \cellcolor[rgb]{0.551,0.745,0.605} 1070.13 \\
 & 500M & \cellcolor[rgb]{0.685,0.821,0.723} 5.80 & \cellcolor[rgb]{0.949,0.971,0.955} 79.71 & --- & --- & \cellcolor[rgb]{1.000,1.000,1.000} 79.99 & \cellcolor[rgb]{0.979,0.988,0.981} 180.70 & \cellcolor[rgb]{1.000,1.000,1.000} 75.14 & \cellcolor[rgb]{0.993,0.996,0.994} 316.46 \\
 & 1.5B & \cellcolor[rgb]{1.000,1.000,1.000} 5.61 & \cellcolor[rgb]{0.991,0.983,0.983} 73.83 & --- & --- & \cellcolor[rgb]{0.834,0.906,0.854} 83.29 & \cellcolor[rgb]{0.862,0.758,0.758} 93.31 & \cellcolor[rgb]{0.940,0.966,0.947} 76.07 & \cellcolor[rgb]{0.866,0.765,0.765} 193.02 \\
\midrule
\multirow[c]{3}{*}{10 Tests} & MV & \cellcolor[rgb]{0.774,0.871,0.801} 5.73 & \cellcolor[rgb]{0.536,0.737,0.592} 122.34 & --- & --- & \cellcolor[rgb]{0.510,0.722,0.569} 88.60 & \cellcolor[rgb]{0.681,0.819,0.720} 571.02 & \cellcolor[rgb]{0.566,0.754,0.618} 81.66 & \cellcolor[rgb]{0.736,0.850,0.767} 511.41 \\
 & 500M & \cellcolor[rgb]{0.510,0.722,0.569} 5.95 & \cellcolor[rgb]{0.953,0.974,0.959} 79.18 & --- & --- & \cellcolor[rgb]{0.652,0.802,0.693} 86.44 & \cellcolor[rgb]{0.990,0.982,0.982} 157.04 & \cellcolor[rgb]{0.520,0.727,0.577} 82.33 & \cellcolor[rgb]{0.863,0.758,0.758} 189.80 \\
 & 1.5B & \cellcolor[rgb]{0.743,0.854,0.774} 5.75 & \cellcolor[rgb]{0.978,0.962,0.962} 73.53 & --- & --- & \cellcolor[rgb]{0.539,0.738,0.595} 88.15 & \cellcolor[rgb]{0.856,0.747,0.747} 91.79 & \cellcolor[rgb]{0.510,0.722,0.569} 82.47 & \cellcolor[rgb]{0.722,0.510,0.510} 142.11 \\
\midrule
\multicolumn{2}{c|}{$\strongestVerifier{}$} & \cellcolor[rgb]{0.510,0.722,0.569} 7.83 & \cellcolor[rgb]{0.722,0.510,0.510} 2.23 & \cellcolor[rgb]{0.510,0.722,0.569} 92.72 & \cellcolor[rgb]{0.722,0.510,0.510} 173.71 & \cellcolor[rgb]{0.510,0.722,0.569} 94.06 & \cellcolor[rgb]{0.722,0.510,0.510} 20.09 & \cellcolor[rgb]{0.510,0.722,0.569} 85.64 & \cellcolor[rgb]{0.722,0.510,0.510} 20.97 \\
\bottomrule
\end{tabular}

    }
\end{table*}

%% file: tables/scale/results_table_3B_1_0.tex
\begin{table*}
    \centering
    
    \caption{Results for the generation with Qwen 2.5 Coder 3B instruct with $T=1.0$, $n=128$, $top_p=0.95$, and 1024 tokens.  {\colorbox{textgreen}{Green backgrounds}} is higher performance while {\colorbox{textred}{Red backgrounds}} is lower performance with respect to the entire column. If $\verifier{}$ is ``---'' that means no pruning is done. $\strongestVerifier{}$ is the case where \emph{all} test cases are run.}\label{tab:3B-1.0-overall-results}
    \resizebox{0.9\textwidth}{!}{%
    \begin{tabular}{ll|cc|cc|cc|cc}
\toprule
 &  & \multicolumn{2}{c|}{CodeContests} & \multicolumn{2}{c|}{GSM8K} & \multicolumn{2}{c|}{HumanEval} & \multicolumn{2}{c}{MBPP} \\ Filter & Ranker & Bof64 & PPS & Bof64 & PPS & Bof64 & PPS & Bof64 & PPS \\
\midrule
\multirow[c]{2}{*}{---} & 500M & \cellcolor[rgb]{0.722,0.510,0.510} 4.52 & \cellcolor[rgb]{0.993,0.987,0.987} 64.02 & \cellcolor[rgb]{0.722,0.510,0.510} 75.48 & \cellcolor[rgb]{0.510,0.722,0.569} 339.96 & \cellcolor[rgb]{0.722,0.510,0.510} 75.50 & \cellcolor[rgb]{0.985,0.991,0.987} 159.60 & \cellcolor[rgb]{0.722,0.510,0.510} 60.70 & \cellcolor[rgb]{0.852,0.916,0.870} 387.75 \\
 & 1.5B & \cellcolor[rgb]{0.751,0.562,0.562} 5.76 & \cellcolor[rgb]{0.722,0.510,0.510} 25.14 & \cellcolor[rgb]{0.510,0.722,0.569} 81.89 & \cellcolor[rgb]{0.722,0.510,0.510} 134.53 & \cellcolor[rgb]{0.830,0.701,0.701} 79.90 & \cellcolor[rgb]{0.722,0.510,0.510} 65.21 & \cellcolor[rgb]{0.795,0.640,0.640} 67.54 & \cellcolor[rgb]{0.815,0.675,0.675} 159.58 \\
\midrule
\multirow[c]{3}{*}{1 Test} & MV & \cellcolor[rgb]{0.845,0.728,0.728} 9.65 & \cellcolor[rgb]{0.510,0.722,0.569} 122.58 & --- & --- & \cellcolor[rgb]{0.903,0.829,0.829} 82.39 & \cellcolor[rgb]{0.557,0.748,0.610} 1057.72 & \cellcolor[rgb]{0.991,0.984,0.984} 74.99 & \cellcolor[rgb]{0.510,0.722,0.569} 1217.04 \\
 & 500M & \cellcolor[rgb]{0.942,0.898,0.898} 10.93 & \cellcolor[rgb]{0.979,0.988,0.982} 66.69 & --- & --- & \cellcolor[rgb]{0.812,0.668,0.668} 79.15 & \cellcolor[rgb]{1.000,1.000,1.000} 142.54 & \cellcolor[rgb]{0.846,0.729,0.729} 72.22 & \cellcolor[rgb]{1.000,1.000,1.000} 294.73 \\
 & 1.5B & \cellcolor[rgb]{1.000,1.000,1.000} 11.39 & \cellcolor[rgb]{0.848,0.733,0.733} 58.02 & --- & --- & \cellcolor[rgb]{0.957,0.924,0.924} 84.01 & \cellcolor[rgb]{0.753,0.564,0.564} 66.80 & \cellcolor[rgb]{1.000,1.000,1.000} 75.09 & \cellcolor[rgb]{0.848,0.733,0.733} 169.51 \\
\midrule
\multirow[c]{3}{*}{3 Tests} & MV & \cellcolor[rgb]{0.994,0.989,0.989} 11.34 & \cellcolor[rgb]{0.541,0.739,0.596} 118.91 & --- & --- & \cellcolor[rgb]{0.964,0.979,0.968} 85.78 & \cellcolor[rgb]{0.510,0.722,0.569} 1208.37 & \cellcolor[rgb]{0.878,0.931,0.893} 77.72 & \cellcolor[rgb]{0.555,0.747,0.609} 1074.66 \\
 & 500M & \cellcolor[rgb]{0.784,0.877,0.810} 12.40 & \cellcolor[rgb]{0.993,0.988,0.988} 64.03 & --- & --- & \cellcolor[rgb]{1.000,1.000,1.000} 85.29 & \cellcolor[rgb]{0.990,0.994,0.991} 153.96 & \cellcolor[rgb]{0.988,0.979,0.979} 74.96 & \cellcolor[rgb]{0.989,0.994,0.990} 301.64 \\
 & 1.5B & \cellcolor[rgb]{0.612,0.779,0.658} 12.83 & \cellcolor[rgb]{0.850,0.736,0.736} 58.40 & --- & --- & \cellcolor[rgb]{0.784,0.877,0.810} 88.22 & \cellcolor[rgb]{0.862,0.757,0.757} 72.96 & \cellcolor[rgb]{0.891,0.938,0.904} 77.45 & \cellcolor[rgb]{0.865,0.763,0.763} 177.22 \\
\midrule
\multirow[c]{3}{*}{10 Tests} & MV & \cellcolor[rgb]{0.835,0.906,0.855} 12.16 & \cellcolor[rgb]{0.564,0.752,0.616} 116.15 & --- & --- & \cellcolor[rgb]{0.671,0.813,0.711} 89.54 & \cellcolor[rgb]{0.673,0.814,0.712} 684.03 & \cellcolor[rgb]{0.519,0.727,0.576} 84.10 & \cellcolor[rgb]{0.735,0.850,0.767} 509.97 \\
 & 500M & \cellcolor[rgb]{0.689,0.824,0.727} 12.67 & \cellcolor[rgb]{0.945,0.969,0.951} 70.78 & --- & --- & \cellcolor[rgb]{0.719,0.841,0.753} 89.01 & \cellcolor[rgb]{0.990,0.982,0.982} 137.37 & \cellcolor[rgb]{0.586,0.765,0.636} 83.04 & \cellcolor[rgb]{0.866,0.765,0.765} 178.20 \\
 & 1.5B & \cellcolor[rgb]{0.510,0.722,0.569} 13.04 & \cellcolor[rgb]{1.000,1.000,1.000} 64.18 & --- & --- & \cellcolor[rgb]{0.510,0.722,0.569} 91.33 & \cellcolor[rgb]{0.849,0.735,0.735} 71.77 & \cellcolor[rgb]{0.510,0.722,0.569} 84.24 & \cellcolor[rgb]{0.722,0.510,0.510} 131.16 \\
\midrule
\multicolumn{2}{c|}{$\strongestVerifier{}$} & \cellcolor[rgb]{0.510,0.722,0.569} 14.92 & \cellcolor[rgb]{0.722,0.510,0.510} 2.54 & \cellcolor[rgb]{0.510,0.722,0.569} 96.63 & \cellcolor[rgb]{0.722,0.510,0.510} 647.01 & \cellcolor[rgb]{0.510,0.722,0.569} 95.41 & \cellcolor[rgb]{0.722,0.510,0.510} 22.13 & \cellcolor[rgb]{0.510,0.722,0.569} 88.23 & \cellcolor[rgb]{0.722,0.510,0.510} 12.48 \\
\bottomrule
\end{tabular}

    }
\end{table*}

%% file: tables/scale/results_table_14B_1_0.tex
\begin{table*}
    \centering
    
    \caption{Results for the generation with Qwen 2.5 Coder 14B instruct with $T=1.0$, $n=128$, $top_p=0.95$, and 1024 tokens.  {\colorbox{textgreen}{Green backgrounds}} is higher performance while {\colorbox{textred}{Red backgrounds}} is lower performance with respect to the entire column. If $\verifier{}$ is ``---'' that means no pruning is done. $\strongestVerifier{}$ is the case where \emph{all} test cases are run.}\label{tab:14B-1.0-overall-results}
    \resizebox{0.9\textwidth}{!}{%
    \begin{tabular}{ll|cc|cc|cc|cc}
\toprule
 &  & \multicolumn{2}{c|}{CodeContests} & \multicolumn{2}{c|}{GSM8K} & \multicolumn{2}{c|}{HumanEval} & \multicolumn{2}{c}{MBPP} \\ Filter & Ranker & Bof64 & PPS & Bof64 & PPS & Bof64 & PPS & Bof64 & PPS \\
\midrule
\multirow[c]{2}{*}{---} & 500M & \cellcolor[rgb]{0.722,0.510,0.510} 10.99 & \cellcolor[rgb]{1.000,1.000,1.000} 59.83 & \cellcolor[rgb]{0.722,0.510,0.510} 90.53 & \cellcolor[rgb]{0.510,0.722,0.569} 323.74 & \cellcolor[rgb]{0.722,0.510,0.510} 83.23 & \cellcolor[rgb]{0.974,0.985,0.977} 172.96 & \cellcolor[rgb]{0.722,0.510,0.510} 68.85 & \cellcolor[rgb]{0.790,0.881,0.815} 328.09 \\
 & 1.5B & \cellcolor[rgb]{0.742,0.546,0.546} 12.86 & \cellcolor[rgb]{0.722,0.510,0.510} 23.52 & \cellcolor[rgb]{0.510,0.722,0.569} 91.49 & \cellcolor[rgb]{0.722,0.510,0.510} 127.83 & \cellcolor[rgb]{0.826,0.694,0.694} 85.44 & \cellcolor[rgb]{0.861,0.756,0.756} 70.92 & \cellcolor[rgb]{0.806,0.659,0.659} 73.28 & \cellcolor[rgb]{0.835,0.710,0.710} 137.09 \\
\midrule
\multirow[c]{3}{*}{1 Test} & MV & \cellcolor[rgb]{0.852,0.740,0.740} 22.76 & \cellcolor[rgb]{0.510,0.722,0.569} 132.92 & --- & --- & \cellcolor[rgb]{1.000,1.000,1.000} 88.67 & \cellcolor[rgb]{0.510,0.722,0.569} 1197.05 & \cellcolor[rgb]{0.944,0.968,0.951} 79.82 & \cellcolor[rgb]{0.510,0.722,0.569} 1360.04 \\
 & 500M & \cellcolor[rgb]{0.926,0.870,0.870} 24.47 & \cellcolor[rgb]{0.954,0.974,0.960} 66.16 & --- & --- & \cellcolor[rgb]{0.809,0.664,0.664} 85.08 & \cellcolor[rgb]{1.000,1.000,1.000} 143.42 & \cellcolor[rgb]{0.844,0.726,0.726} 75.26 & \cellcolor[rgb]{1.000,1.000,1.000} 264.22 \\
 & 1.5B & \cellcolor[rgb]{1.000,1.000,1.000} 25.52 & \cellcolor[rgb]{0.850,0.735,0.735} 52.46 & --- & --- & \cellcolor[rgb]{0.901,0.826,0.826} 86.89 & \cellcolor[rgb]{0.732,0.528,0.528} 67.51 & \cellcolor[rgb]{0.925,0.867,0.867} 77.29 & \cellcolor[rgb]{0.849,0.733,0.733} 139.90 \\
\midrule
\multirow[c]{3}{*}{3 Tests} & MV & \cellcolor[rgb]{0.916,0.851,0.851} 24.32 & \cellcolor[rgb]{0.546,0.742,0.601} 127.07 & --- & --- & \cellcolor[rgb]{0.763,0.865,0.791} 90.36 & \cellcolor[rgb]{0.531,0.733,0.587} 1130.49 & \cellcolor[rgb]{0.856,0.918,0.873} 81.65 & \cellcolor[rgb]{0.566,0.753,0.618} 1127.35 \\
 & 500M & \cellcolor[rgb]{0.835,0.906,0.854} 26.07 & \cellcolor[rgb]{0.979,0.962,0.962} 59.08 & --- & --- & \cellcolor[rgb]{0.968,0.944,0.944} 88.10 & \cellcolor[rgb]{0.995,0.997,0.996} 148.71 & \cellcolor[rgb]{0.909,0.839,0.839} 77.00 & \cellcolor[rgb]{0.989,0.994,0.991} 267.42 \\
 & 1.5B & \cellcolor[rgb]{0.718,0.840,0.752} 26.60 & \cellcolor[rgb]{0.841,0.720,0.720} 50.44 & --- & --- & \cellcolor[rgb]{0.956,0.975,0.962} 88.98 & \cellcolor[rgb]{0.851,0.738,0.738} 70.44 & \cellcolor[rgb]{1.000,1.000,1.000} 78.66 & \cellcolor[rgb]{0.864,0.760,0.760} 145.13 \\
\midrule
\multirow[c]{3}{*}{10 Tests} & MV & \cellcolor[rgb]{0.907,0.947,0.918} 25.83 & \cellcolor[rgb]{0.588,0.766,0.637} 120.53 & --- & --- & \cellcolor[rgb]{0.510,0.722,0.569} 91.62 & \cellcolor[rgb]{0.678,0.817,0.717} 662.39 & \cellcolor[rgb]{0.510,0.722,0.569} 87.72 & \cellcolor[rgb]{0.752,0.859,0.782} 349.60 \\
 & 500M & \cellcolor[rgb]{0.521,0.728,0.578} 28.02 & \cellcolor[rgb]{0.947,0.970,0.954} 67.14 & --- & --- & \cellcolor[rgb]{0.744,0.854,0.774} 90.47 & \cellcolor[rgb]{0.971,0.948,0.948} 128.04 & \cellcolor[rgb]{0.626,0.788,0.671} 85.83 & \cellcolor[rgb]{0.878,0.785,0.785} 157.48 \\
 & 1.5B & \cellcolor[rgb]{0.510,0.722,0.569} 28.10 & \cellcolor[rgb]{0.932,0.881,0.881} 57.46 & --- & --- & \cellcolor[rgb]{0.660,0.807,0.701} 90.88 & \cellcolor[rgb]{0.722,0.510,0.510} 67.25 & \cellcolor[rgb]{0.587,0.765,0.636} 86.47 & \cellcolor[rgb]{0.722,0.510,0.510} 112.77 \\
\midrule
\multicolumn{2}{c|}{$\strongestVerifier{}$} & \cellcolor[rgb]{0.510,0.722,0.569} 31.98 & \cellcolor[rgb]{0.722,0.510,0.510} 3.65 & \cellcolor[rgb]{0.510,0.722,0.569} 98.61 & \cellcolor[rgb]{0.722,0.510,0.510} 153.10 & \cellcolor[rgb]{0.510,0.722,0.569} 96.57 & \cellcolor[rgb]{0.722,0.510,0.510} 24.43 & \cellcolor[rgb]{0.510,0.722,0.569} 90.29 & \cellcolor[rgb]{0.722,0.510,0.510} 17.11 \\
\bottomrule
\end{tabular}

    }
\end{table*}

%% file: tables/scale/results_table_7B_0_2.tex
\begin{table*}
    \centering
    
    \caption{Results for the generation with Qwen 2.5 Coder 7B instruct with $T=0.2$, $n=128$, $top_p=0.95$, and 1024 tokens.  {\colorbox{textgreen}{Green backgrounds}} is higher performance while {\colorbox{textred}{Red backgrounds}} is lower performance with respect to the entire column. If $\verifier{}$ is ``---'' that means no pruning is done. $\strongestVerifier{}$ is the case where \emph{all} test cases are run.}\label{tab:7B-0.2-overall-results}
    \resizebox{0.9\textwidth}{!}{%
    \begin{tabular}{ll|cc|cc|cc|cc}
\toprule
 &  & \multicolumn{2}{c|}{CodeContests} & \multicolumn{2}{c|}{GSM8K} & \multicolumn{2}{c|}{HumanEval} & \multicolumn{2}{c}{MBPP} \\ Filter & Ranker & Bof64 & PPS & Bof64 & PPS & Bof64 & PPS & Bof64 & PPS \\
\midrule
\multirow[c]{2}{*}{---} & 500M & \cellcolor[rgb]{0.734,0.531,0.531} 5.86 & \cellcolor[rgb]{0.827,0.696,0.696} 61.72 & \cellcolor[rgb]{0.722,0.510,0.510} 84.94 & \cellcolor[rgb]{0.510,0.722,0.569} 259.14 & \cellcolor[rgb]{0.722,0.510,0.510} 85.11 & \cellcolor[rgb]{0.985,0.992,0.987} 131.74 & \cellcolor[rgb]{0.722,0.510,0.510} 71.26 & \cellcolor[rgb]{0.891,0.938,0.904} 258.15 \\
 & 1.5B & \cellcolor[rgb]{0.722,0.510,0.510} 5.38 & \cellcolor[rgb]{0.722,0.510,0.510} 24.54 & \cellcolor[rgb]{0.510,0.722,0.569} 86.11 & \cellcolor[rgb]{0.722,0.510,0.510} 110.67 & \cellcolor[rgb]{0.809,0.664,0.664} 86.44 & \cellcolor[rgb]{0.796,0.641,0.641} 57.42 & \cellcolor[rgb]{0.784,0.620,0.620} 73.93 & \cellcolor[rgb]{0.776,0.605,0.605} 115.38 \\
\midrule
\multirow[c]{3}{*}{1 Test} & MV & \cellcolor[rgb]{0.860,0.754,0.754} 10.82 & \cellcolor[rgb]{0.510,0.722,0.569} 173.82 & --- & --- & \cellcolor[rgb]{1.000,1.000,1.000} 88.41 & \cellcolor[rgb]{0.579,0.761,0.630} 961.83 & \cellcolor[rgb]{1.000,1.000,1.000} 78.24 & \cellcolor[rgb]{0.745,0.855,0.776} 316.36 \\
 & 500M & \cellcolor[rgb]{0.864,0.760,0.760} 10.86 & \cellcolor[rgb]{1.000,1.000,1.000} 81.39 & --- & --- & \cellcolor[rgb]{0.788,0.627,0.627} 86.12 & \cellcolor[rgb]{1.000,1.000,1.000} 116.61 & \cellcolor[rgb]{0.856,0.746,0.746} 76.98 & \cellcolor[rgb]{1.000,1.000,1.000} 234.99 \\
 & 1.5B & \cellcolor[rgb]{1.000,1.000,1.000} 11.72 & \cellcolor[rgb]{0.842,0.721,0.721} 66.73 & --- & --- & \cellcolor[rgb]{0.953,0.917,0.917} 88.01 & \cellcolor[rgb]{0.722,0.510,0.510} 56.21 & \cellcolor[rgb]{0.890,0.807,0.807} 77.42 & \cellcolor[rgb]{0.843,0.723,0.723} 128.61 \\
\midrule
\multirow[c]{3}{*}{3 Tests} & MV & \cellcolor[rgb]{0.995,0.992,0.992} 11.69 & \cellcolor[rgb]{0.563,0.752,0.615} 163.05 & --- & --- & \cellcolor[rgb]{0.744,0.855,0.775} 89.63 & \cellcolor[rgb]{0.510,0.722,0.569} 1196.20 & \cellcolor[rgb]{0.834,0.906,0.854} 79.63 & \cellcolor[rgb]{0.510,0.722,0.569} 1010.78 \\
 & 500M & \cellcolor[rgb]{0.813,0.894,0.835} 12.11 & \cellcolor[rgb]{0.947,0.970,0.953} 90.59 & --- & --- & \cellcolor[rgb]{0.962,0.934,0.934} 88.09 & \cellcolor[rgb]{0.996,0.998,0.996} 120.93 & \cellcolor[rgb]{0.905,0.833,0.833} 77.53 & \cellcolor[rgb]{0.973,0.985,0.976} 240.79 \\
 & 1.5B & \cellcolor[rgb]{0.539,0.738,0.594} 12.38 & \cellcolor[rgb]{0.989,0.981,0.981} 80.77 & --- & --- & \cellcolor[rgb]{0.766,0.867,0.794} 89.52 & \cellcolor[rgb]{0.863,0.759,0.759} 59.52 & \cellcolor[rgb]{0.992,0.995,0.993} 78.31 & \cellcolor[rgb]{0.866,0.763,0.763} 135.70 \\
\midrule
\multirow[c]{3}{*}{10 Tests} & MV & \cellcolor[rgb]{0.822,0.899,0.844} 12.09 & \cellcolor[rgb]{0.592,0.768,0.641} 157.14 & --- & --- & \cellcolor[rgb]{0.510,0.722,0.569} 90.85 & \cellcolor[rgb]{0.684,0.820,0.722} 611.21 & \cellcolor[rgb]{0.510,0.722,0.569} 81.60 & \cellcolor[rgb]{0.726,0.844,0.759} 373.46 \\
 & 500M & \cellcolor[rgb]{0.582,0.763,0.632} 12.35 & \cellcolor[rgb]{0.959,0.977,0.964} 88.55 & --- & --- & \cellcolor[rgb]{0.821,0.898,0.843} 89.26 & \cellcolor[rgb]{0.971,0.950,0.950} 104.69 & \cellcolor[rgb]{0.631,0.791,0.676} 80.95 & \cellcolor[rgb]{0.877,0.783,0.783} 143.88 \\
 & 1.5B & \cellcolor[rgb]{0.510,0.722,0.569} 12.40 & \cellcolor[rgb]{0.978,0.961,0.961} 80.12 & --- & --- & \cellcolor[rgb]{0.600,0.773,0.648} 90.38 & \cellcolor[rgb]{0.731,0.526,0.526} 56.36 & \cellcolor[rgb]{0.564,0.752,0.616} 81.31 & \cellcolor[rgb]{0.722,0.510,0.510} 104.77 \\
\midrule
\multicolumn{2}{c|}{$\strongestVerifier{}$} & \cellcolor[rgb]{0.510,0.722,0.569} 12.40 & \cellcolor[rgb]{0.722,0.510,0.510} 3.09 & \cellcolor[rgb]{0.510,0.722,0.569} 91.17 & \cellcolor[rgb]{0.722,0.510,0.510} 504.11 & \cellcolor[rgb]{0.510,0.722,0.569} 91.46 & \cellcolor[rgb]{0.722,0.510,0.510} 11.63 & \cellcolor[rgb]{0.510,0.722,0.569} 82.40 & \cellcolor[rgb]{0.722,0.510,0.510} 6.97 \\
\bottomrule
\end{tabular}

    }
\end{table*}

%% file: tables/scale/results_table_7B_0_4.tex
\begin{table*}
    \centering
    
    \caption{Results for the generation with Qwen 2.5 Coder 7B instruct with $T=0.4$, $n=128$, $top_p=0.95$, and 1024 tokens.  {\colorbox{textgreen}{Green backgrounds}} is higher performance while {\colorbox{textred}{Red backgrounds}} is lower performance with respect to the entire column. If $\verifier{}$ is ``---'' that means no pruning is done. $\strongestVerifier{}$ is the case where \emph{all} test cases are run.}\label{tab:7B-0.4-overall-results}
    \resizebox{0.9\textwidth}{!}{%
    \begin{tabular}{ll|cc|cc|cc|cc}
\toprule
 &  & \multicolumn{2}{c|}{CodeContests} & \multicolumn{2}{c|}{GSM8K} & \multicolumn{2}{c|}{HumanEval} & \multicolumn{2}{c}{MBPP} \\ Filter & Ranker & Bof64 & PPS & Bof64 & PPS & Bof64 & PPS & Bof64 & PPS \\
\midrule
\multirow[c]{2}{*}{---} & 500M & \cellcolor[rgb]{0.722,0.510,0.510} 5.61 & \cellcolor[rgb]{0.854,0.742,0.742} 65.65 & \cellcolor[rgb]{0.722,0.510,0.510} 86.15 & \cellcolor[rgb]{0.510,0.722,0.569} 299.19 & \cellcolor[rgb]{0.722,0.510,0.510} 82.74 & \cellcolor[rgb]{0.984,0.991,0.986} 151.34 & \cellcolor[rgb]{0.722,0.510,0.510} 69.13 & \cellcolor[rgb]{0.816,0.895,0.838} 317.32 \\
 & 1.5B & \cellcolor[rgb]{0.750,0.561,0.561} 7.29 & \cellcolor[rgb]{0.722,0.510,0.510} 26.06 & \cellcolor[rgb]{0.510,0.722,0.569} 88.40 & \cellcolor[rgb]{0.722,0.510,0.510} 120.73 & \cellcolor[rgb]{0.828,0.697,0.697} 85.35 & \cellcolor[rgb]{0.816,0.676,0.676} 63.50 & \cellcolor[rgb]{0.792,0.634,0.634} 72.65 & \cellcolor[rgb]{0.829,0.699,0.699} 134.96 \\
\midrule
\multirow[c]{3}{*}{1 Test} & MV & \cellcolor[rgb]{0.851,0.737,0.737} 13.11 & \cellcolor[rgb]{0.510,0.722,0.569} 163.12 & --- & --- & \cellcolor[rgb]{1.000,1.000,1.000} 87.80 & \cellcolor[rgb]{0.510,0.722,0.569} 1146.83 & \cellcolor[rgb]{0.942,0.967,0.949} 79.15 & \cellcolor[rgb]{0.510,0.722,0.569} 1257.67 \\
 & 500M & \cellcolor[rgb]{0.906,0.835,0.835} 14.29 & \cellcolor[rgb]{1.000,1.000,1.000} 77.29 & --- & --- & \cellcolor[rgb]{0.787,0.625,0.625} 84.35 & \cellcolor[rgb]{1.000,1.000,1.000} 134.12 & \cellcolor[rgb]{0.847,0.731,0.731} 75.38 & \cellcolor[rgb]{1.000,1.000,1.000} 254.88 \\
 & 1.5B & \cellcolor[rgb]{0.954,0.920,0.920} 14.91 & \cellcolor[rgb]{0.848,0.733,0.733} 64.04 & --- & --- & \cellcolor[rgb]{0.929,0.875,0.875} 86.96 & \cellcolor[rgb]{0.779,0.612,0.612} 62.73 & \cellcolor[rgb]{0.901,0.826,0.826} 76.75 & \cellcolor[rgb]{0.845,0.728,0.728} 138.36 \\
\midrule
\multirow[c]{3}{*}{3 Tests} & MV & \cellcolor[rgb]{1.000,1.000,1.000} 15.50 & \cellcolor[rgb]{0.558,0.749,0.611} 153.76 & --- & --- & \cellcolor[rgb]{0.729,0.846,0.761} 89.56 & \cellcolor[rgb]{0.511,0.722,0.569} 1143.86 & \cellcolor[rgb]{0.829,0.903,0.849} 80.52 & \cellcolor[rgb]{0.523,0.729,0.580} 1207.37 \\
 & 500M & \cellcolor[rgb]{0.924,0.957,0.933} 15.64 & \cellcolor[rgb]{0.994,0.997,0.995} 78.18 & --- & --- & \cellcolor[rgb]{0.952,0.915,0.915} 87.23 & \cellcolor[rgb]{0.996,0.998,0.997} 138.00 & \cellcolor[rgb]{0.905,0.833,0.833} 76.82 & \cellcolor[rgb]{0.981,0.989,0.983} 261.47 \\
 & 1.5B & \cellcolor[rgb]{0.589,0.767,0.638} 16.18 & \cellcolor[rgb]{0.892,0.810,0.810} 69.93 & --- & --- & \cellcolor[rgb]{0.841,0.909,0.860} 88.85 & \cellcolor[rgb]{0.863,0.758,0.758} 65.37 & \cellcolor[rgb]{1.000,1.000,1.000} 78.44 & \cellcolor[rgb]{0.865,0.762,0.762} 144.88 \\
\midrule
\multirow[c]{3}{*}{10 Tests} & MV & \cellcolor[rgb]{0.782,0.876,0.808} 15.90 & \cellcolor[rgb]{0.587,0.765,0.637} 147.99 & --- & --- & \cellcolor[rgb]{0.510,0.722,0.569} 90.78 & \cellcolor[rgb]{0.675,0.816,0.714} 639.57 & \cellcolor[rgb]{0.510,0.722,0.569} 83.29 & \cellcolor[rgb]{0.749,0.858,0.779} 358.78 \\
 & 500M & \cellcolor[rgb]{0.719,0.840,0.753} 16.00 & \cellcolor[rgb]{0.968,0.982,0.972} 82.17 & --- & --- & \cellcolor[rgb]{0.777,0.873,0.804} 89.27 & \cellcolor[rgb]{0.965,0.938,0.938} 116.53 & \cellcolor[rgb]{0.637,0.794,0.681} 82.32 & \cellcolor[rgb]{0.880,0.789,0.789} 157.60 \\
 & 1.5B & \cellcolor[rgb]{0.510,0.722,0.569} 16.29 & \cellcolor[rgb]{0.949,0.911,0.911} 73.83 & --- & --- & \cellcolor[rgb]{0.644,0.798,0.687} 90.03 & \cellcolor[rgb]{0.722,0.510,0.510} 61.52 & \cellcolor[rgb]{0.535,0.736,0.591} 83.10 & \cellcolor[rgb]{0.722,0.510,0.510} 112.16 \\
\midrule
\multicolumn{2}{c|}{$\strongestVerifier{}$} & \cellcolor[rgb]{0.510,0.722,0.569} 16.74 & \cellcolor[rgb]{0.722,0.510,0.510} 3.83 & \cellcolor[rgb]{0.510,0.722,0.569} 95.21 & \cellcolor[rgb]{0.722,0.510,0.510} 60.02 & \cellcolor[rgb]{0.510,0.722,0.569} 92.57 & \cellcolor[rgb]{0.722,0.510,0.510} 15.11 & \cellcolor[rgb]{0.510,0.722,0.569} 84.55 & \cellcolor[rgb]{0.722,0.510,0.510} 14.80 \\
\bottomrule
\end{tabular}

    }
\end{table*}

%% file: tables/scale/results_table_7B_0_6.tex
\begin{table*}
    \centering
    
    \caption{Results for the generation with Qwen 2.5 Coder 7B instruct with $T=0.6$, $n=128$, $top_p=0.95$, and 1024 tokens.  {\colorbox{textgreen}{Green backgrounds}} is higher performance while {\colorbox{textred}{Red backgrounds}} is lower performance with respect to the entire column. If $\verifier{}$ is ``---'' that means no pruning is done. $\strongestVerifier{}$ is the case where \emph{all} test cases are run.}\label{tab:7B-0.6-overall-results}
    \resizebox{0.9\textwidth}{!}{%
    \begin{tabular}{ll|cc|cc|cc|cc}
\toprule
 &  & \multicolumn{2}{c|}{CodeContests} & \multicolumn{2}{c|}{GSM8K} & \multicolumn{2}{c|}{HumanEval} & \multicolumn{2}{c}{MBPP} \\ Filter & Ranker & Bof64 & PPS & Bof64 & PPS & Bof64 & PPS & Bof64 & PPS \\
\midrule
\multirow[c]{2}{*}{---} & 500M & \cellcolor[rgb]{0.722,0.510,0.510} 6.00 & \cellcolor[rgb]{0.855,0.745,0.745} 66.83 & \cellcolor[rgb]{0.722,0.510,0.510} 85.73 & \cellcolor[rgb]{0.510,0.722,0.569} 317.18 & \cellcolor[rgb]{0.722,0.510,0.510} 83.09 & \cellcolor[rgb]{0.971,0.984,0.975} 163.23 & \cellcolor[rgb]{0.722,0.510,0.510} 67.14 & \cellcolor[rgb]{0.769,0.869,0.797} 351.91 \\
 & 1.5B & \cellcolor[rgb]{0.745,0.551,0.551} 7.21 & \cellcolor[rgb]{0.722,0.510,0.510} 26.53 & \cellcolor[rgb]{0.510,0.722,0.569} 88.19 & \cellcolor[rgb]{0.722,0.510,0.510} 127.31 & \cellcolor[rgb]{0.811,0.668,0.668} 84.78 & \cellcolor[rgb]{0.784,0.620,0.620} 67.77 & \cellcolor[rgb]{0.791,0.632,0.632} 71.45 & \cellcolor[rgb]{0.854,0.744,0.744} 146.91 \\
\midrule
\multirow[c]{3}{*}{1 Test} & MV & \cellcolor[rgb]{0.856,0.747,0.747} 12.99 & \cellcolor[rgb]{0.510,0.722,0.569} 145.28 & --- & --- & \cellcolor[rgb]{0.933,0.882,0.882} 86.77 & \cellcolor[rgb]{0.510,0.722,0.569} 1280.37 & \cellcolor[rgb]{1.000,1.000,1.000} 77.83 & \cellcolor[rgb]{0.510,0.722,0.569} 1183.06 \\
 & 500M & \cellcolor[rgb]{0.881,0.791,0.791} 13.46 & \cellcolor[rgb]{1.000,1.000,1.000} 74.06 & --- & --- & \cellcolor[rgb]{0.816,0.676,0.676} 84.87 & \cellcolor[rgb]{1.000,1.000,1.000} 140.68 & \cellcolor[rgb]{0.845,0.727,0.727} 74.82 & \cellcolor[rgb]{1.000,1.000,1.000} 264.62 \\
 & 1.5B & \cellcolor[rgb]{0.961,0.978,0.965} 14.94 & \cellcolor[rgb]{0.839,0.717,0.717} 62.05 & --- & --- & \cellcolor[rgb]{0.918,0.856,0.856} 86.55 & \cellcolor[rgb]{0.727,0.520,0.520} 66.63 & \cellcolor[rgb]{0.944,0.901,0.901} 77.02 & \cellcolor[rgb]{0.844,0.726,0.726} 144.47 \\
\midrule
\multirow[c]{3}{*}{3 Tests} & MV & \cellcolor[rgb]{0.993,0.988,0.988} 14.76 & \cellcolor[rgb]{0.529,0.732,0.585} 142.47 & --- & --- & \cellcolor[rgb]{0.735,0.850,0.767} 89.71 & \cellcolor[rgb]{0.522,0.729,0.580} 1232.14 & \cellcolor[rgb]{0.871,0.927,0.887} 80.06 & \cellcolor[rgb]{0.561,0.750,0.613} 1011.88 \\
 & 500M & \cellcolor[rgb]{1.000,1.000,1.000} 14.84 & \cellcolor[rgb]{0.966,0.981,0.970} 78.82 & --- & --- & \cellcolor[rgb]{1.000,1.000,1.000} 87.75 & \cellcolor[rgb]{0.988,0.993,0.989} 150.36 & \cellcolor[rgb]{0.930,0.877,0.877} 76.82 & \cellcolor[rgb]{0.990,0.994,0.991} 268.41 \\
 & 1.5B & \cellcolor[rgb]{0.616,0.782,0.662} 15.82 & \cellcolor[rgb]{0.902,0.827,0.827} 70.11 & --- & --- & \cellcolor[rgb]{0.901,0.944,0.913} 88.50 & \cellcolor[rgb]{0.864,0.760,0.760} 70.82 & \cellcolor[rgb]{0.938,0.965,0.945} 78.91 & \cellcolor[rgb]{0.863,0.758,0.758} 150.06 \\
\midrule
\multirow[c]{3}{*}{10 Tests} & MV & \cellcolor[rgb]{0.690,0.824,0.727} 15.63 & \cellcolor[rgb]{0.570,0.755,0.621} 136.34 & --- & --- & \cellcolor[rgb]{0.510,0.722,0.569} 90.93 & \cellcolor[rgb]{0.711,0.836,0.745} 504.65 & \cellcolor[rgb]{0.510,0.722,0.569} 85.36 & \cellcolor[rgb]{0.753,0.860,0.783} 362.63 \\
 & 500M & \cellcolor[rgb]{0.820,0.898,0.841} 15.30 & \cellcolor[rgb]{0.952,0.973,0.958} 80.81 & --- & --- & \cellcolor[rgb]{0.766,0.867,0.794} 89.52 & \cellcolor[rgb]{0.970,0.947,0.947} 125.36 & \cellcolor[rgb]{0.605,0.775,0.652} 84.09 & \cellcolor[rgb]{0.874,0.778,0.778} 159.63 \\
 & 1.5B & \cellcolor[rgb]{0.510,0.722,0.569} 16.09 & \cellcolor[rgb]{0.962,0.934,0.934} 72.55 & --- & --- & \cellcolor[rgb]{0.739,0.852,0.770} 89.69 & \cellcolor[rgb]{0.722,0.510,0.510} 66.52 & \cellcolor[rgb]{0.514,0.724,0.573} 85.30 & \cellcolor[rgb]{0.722,0.510,0.510} 114.18 \\
\midrule
\multicolumn{2}{c|}{$\strongestVerifier{}$} & \cellcolor[rgb]{0.510,0.722,0.569} 17.65 & \cellcolor[rgb]{0.722,0.510,0.510} 2.13 & \cellcolor[rgb]{0.510,0.722,0.569} 96.40 & \cellcolor[rgb]{0.722,0.510,0.510} 590.56 & \cellcolor[rgb]{0.510,0.722,0.569} 94.05 & \cellcolor[rgb]{0.722,0.510,0.510} 18.06 & \cellcolor[rgb]{0.510,0.722,0.569} 86.23 & \cellcolor[rgb]{0.722,0.510,0.510} 26.87 \\
\bottomrule
\end{tabular}

    }
\end{table*}

%% file: tables/scale/results_table_7B_0_8.tex
\begin{table*}
    \centering
    
    \caption{Results for the generation with Qwen 2.5 Coder 7B instruct with $T=0.8$, $n=128$, $top_p=0.95$, and 1024 tokens.  {\colorbox{textgreen}{Green backgrounds}} is higher performance while {\colorbox{textred}{Red backgrounds}} is lower performance with respect to the entire column. If $\verifier{}$ is ``---'' that means no pruning is done. $\strongestVerifier{}$ is the case where \emph{all} test cases are run.}\label{tab:7B-0.8-overall-results}
    \resizebox{0.9\textwidth}{!}{%
    \begin{tabular}{ll|cc|cc|cc|cc}
\toprule
 &  & \multicolumn{2}{c|}{CodeContests} & \multicolumn{2}{c|}{GSM8K} & \multicolumn{2}{c|}{HumanEval} & \multicolumn{2}{c}{MBPP} \\ Filter & Ranker & Bof64 & PPS & Bof64 & PPS & Bof64 & PPS & Bof64 & PPS \\
\midrule
\multirow[c]{2}{*}{---} & 500M & \cellcolor[rgb]{0.722,0.510,0.510} 5.44 & \cellcolor[rgb]{0.997,0.995,0.995} 67.12 & \cellcolor[rgb]{0.722,0.510,0.510} 85.26 & \cellcolor[rgb]{0.510,0.722,0.569} 328.44 & \cellcolor[rgb]{0.722,0.510,0.510} 80.57 & \cellcolor[rgb]{0.975,0.986,0.978} 172.06 & \cellcolor[rgb]{0.722,0.510,0.510} 66.03 & \cellcolor[rgb]{0.826,0.901,0.847} 374.18 \\
 & 1.5B & \cellcolor[rgb]{0.747,0.555,0.555} 6.96 & \cellcolor[rgb]{0.722,0.510,0.510} 25.49 & \cellcolor[rgb]{0.510,0.722,0.569} 87.97 & \cellcolor[rgb]{0.722,0.510,0.510} 125.81 & \cellcolor[rgb]{0.802,0.651,0.651} 82.54 & \cellcolor[rgb]{0.860,0.753,0.753} 71.43 & \cellcolor[rgb]{0.797,0.643,0.643} 71.12 & \cellcolor[rgb]{0.862,0.756,0.756} 151.33 \\
\midrule
\multirow[c]{3}{*}{1 Test} & MV & \cellcolor[rgb]{0.857,0.748,0.748} 13.54 & \cellcolor[rgb]{0.510,0.722,0.569} 137.85 & --- & --- & \cellcolor[rgb]{0.963,0.979,0.968} 87.43 & \cellcolor[rgb]{0.510,0.722,0.569} 1327.06 & \cellcolor[rgb]{0.908,0.837,0.837} 76.03 & \cellcolor[rgb]{0.510,0.722,0.569} 1097.09 \\
 & 500M & \cellcolor[rgb]{0.899,0.822,0.822} 13.98 & \cellcolor[rgb]{0.976,0.986,0.979} 70.55 & --- & --- & \cellcolor[rgb]{0.839,0.716,0.716} 83.44 & \cellcolor[rgb]{1.000,1.000,1.000} 143.43 & \cellcolor[rgb]{0.852,0.739,0.739} 74.78 & \cellcolor[rgb]{1.000,1.000,1.000} 268.36 \\
 & 1.5B & \cellcolor[rgb]{0.888,0.937,0.902} 14.92 & \cellcolor[rgb]{0.851,0.738,0.738} 59.29 & --- & --- & \cellcolor[rgb]{0.885,0.798,0.798} 84.53 & \cellcolor[rgb]{0.722,0.510,0.510} 67.50 & \cellcolor[rgb]{1.000,1.000,1.000} 77.26 & \cellcolor[rgb]{0.839,0.716,0.716} 145.59 \\
\midrule
\multirow[c]{3}{*}{3 Tests} & MV & \cellcolor[rgb]{0.983,0.969,0.969} 14.46 & \cellcolor[rgb]{0.549,0.744,0.603} 131.96 & --- & --- & \cellcolor[rgb]{0.642,0.797,0.685} 90.72 & \cellcolor[rgb]{0.550,0.745,0.604} 1178.13 & \cellcolor[rgb]{0.896,0.941,0.909} 79.32 & \cellcolor[rgb]{0.522,0.728,0.579} 1063.62 \\
 & 500M & \cellcolor[rgb]{0.789,0.880,0.814} 15.24 & \cellcolor[rgb]{0.926,0.869,0.869} 64.33 & --- & --- & \cellcolor[rgb]{0.967,0.942,0.942} 86.33 & \cellcolor[rgb]{0.993,0.996,0.994} 151.41 & \cellcolor[rgb]{0.957,0.925,0.925} 76.69 & \cellcolor[rgb]{0.997,0.998,0.997} 270.28 \\
 & 1.5B & \cellcolor[rgb]{0.580,0.761,0.630} 15.60 & \cellcolor[rgb]{0.846,0.729,0.729} 57.96 & --- & --- & \cellcolor[rgb]{1.000,1.000,1.000} 87.06 & \cellcolor[rgb]{0.861,0.755,0.755} 71.48 & \cellcolor[rgb]{0.912,0.950,0.923} 79.00 & \cellcolor[rgb]{0.858,0.750,0.750} 150.07 \\
\midrule
\multirow[c]{3}{*}{10 Tests} & MV & \cellcolor[rgb]{1.000,1.000,1.000} 14.56 & \cellcolor[rgb]{0.583,0.763,0.633} 126.77 & --- & --- & \cellcolor[rgb]{0.510,0.722,0.569} 92.10 & \cellcolor[rgb]{0.686,0.822,0.723} 680.40 & \cellcolor[rgb]{0.541,0.740,0.596} 85.77 & \cellcolor[rgb]{0.739,0.852,0.771} 459.95 \\
 & 500M & \cellcolor[rgb]{0.678,0.817,0.717} 15.46 & \cellcolor[rgb]{0.942,0.967,0.949} 75.19 & --- & --- & \cellcolor[rgb]{0.757,0.862,0.786} 89.52 & \cellcolor[rgb]{0.974,0.954,0.954} 129.99 & \cellcolor[rgb]{0.591,0.767,0.640} 84.93 & \cellcolor[rgb]{0.883,0.794,0.794} 169.68 \\
 & 1.5B & \cellcolor[rgb]{0.510,0.722,0.569} 15.70 & \cellcolor[rgb]{1.000,1.000,1.000} 67.22 & --- & --- & \cellcolor[rgb]{0.753,0.860,0.783} 89.56 & \cellcolor[rgb]{0.740,0.543,0.543} 68.03 & \cellcolor[rgb]{0.510,0.722,0.569} 86.31 & \cellcolor[rgb]{0.722,0.510,0.510} 118.63 \\
\midrule
\multicolumn{2}{c|}{$\strongestVerifier{}$} & \cellcolor[rgb]{0.510,0.722,0.569} 17.98 & \cellcolor[rgb]{0.722,0.510,0.510} 3.35 & \cellcolor[rgb]{0.510,0.722,0.569} 97.30 & \cellcolor[rgb]{0.722,0.510,0.510} 582.08 & \cellcolor[rgb]{0.510,0.722,0.569} 95.65 & \cellcolor[rgb]{0.722,0.510,0.510} 19.81 & \cellcolor[rgb]{0.510,0.722,0.569} 89.03 & \cellcolor[rgb]{0.722,0.510,0.510} 19.55 \\
\bottomrule
\end{tabular}

    }
\end{table*}